\documentclass[a4paper,11pt]{article}
\pdfoutput=1
\usepackage{graphicx} 
\usepackage{jheppub}
\usepackage{cleveref}
\usepackage{slashed}
\usepackage{tkz-euclide} 
\usepackage{comment}
\usepackage{subcaption}
\usepackage{amsmath}
\usepackage{mathrsfs}
\usepackage{soul}
\usepackage{enumitem}

\definecolor{darkgreen}{rgb}{0,0.5,0}

\definecolor{c1}{rgb}{0., 0.26, 0.9}
\newcounter{qnumber}

\title{Stability of Superconducting Strings}
\author[1,2,3]{Keisuke Harigaya,}
\author[4]{Xuce Niu,}
\author[4]{Wei Xue,}
\author[4]{and Fengwei Yang}

\affiliation[1]{Department of Physics, University of Chicago,\\
5720 South Ellis Avenue, Chicago, IL 60637, U.S.A.}
\affiliation[2]{Enrico Fermi Institute and Kavli Institute for Cosmological Physics, University of Chicago,\\
933 East 56th Street, Chicago, IL 60637, U.S.A.}
\affiliation[3]{Kavli Institute for the Physics and Mathematics of the Universe (WPI),\\
The University of Tokyo Institutes for Advanced Study, The University of Tokyo,\\
5-1-5 Kashiwanoha, Kashiwa, Chiba 277-8583, Japan}
\affiliation[4]{Institute for Fundamental Theory, Department of Physics,
University of Florida,\\Gainesville, FL 32611, U.S.A. }
\emailAdd{kharigaya@uchicago.edu}
\emailAdd{xuce.niu@ufl.edu}
\emailAdd{weixue@ufl.edu}
\emailAdd{fengwei.yang@ufl.edu}

\begin{document}

\abstract{
We investigate the stability of superconducting strings as bound states of strings and fermion zero modes at both the classical and quantum levels. The dynamics of these superconducting strings can result in a stable configuration, known as a {\it vorton}. We mainly focus on global strings, but the majority of the discussion can be applied to local strings. Using lattice simulations, we study the classical dynamics of superconducting strings and confirm that they relax to the vorton configuration through Nambu-Goldstone boson radiation, with no evidence of over-shooting that would destabilize the vorton. We explore the tunneling of fermion zero modes out of the strings. Both our classical analysis and quantum calculations yield consistent results: the maximum energy of the zero mode significantly exceeds the fermion mass, in contrast to previous literature. Additionally, we introduce a world-sheet formalism to evaluate the decay rate of zero modes into other particles, which constitute the dominant decay channel. We also identify additional processes that trigger zero-mode decay due to non-adiabatic changes of the string configuration. In these decay processes, the rates are suppressed by the curvature of string loops, with exponential suppression for large masses of the final states. We further study the scattering with light charged particles surrounding the string core produced by the zero-mode current and find that a wide zero-mode wavefunction can enhance vorton stability. 
}

\maketitle
\flushbottom

\section{Introduction}

Cosmic strings are topologically stable solutions in field theory that arise when a theory exhibits spontaneous symmetry breaking, 
leading to a non-trivial winding in the field vacuum manifold. 
A simple realization involves a complex scalar field, $\Phi$, with a global $\rm U(1)$ symmetry.  
Cosmic strings form during a phase transition in the early universe via the  
Kibble-Zurek mechanism \cite{Kibble:1976sj, Kibble:1980mv,Zurek:1985qw,Zurek:1996sj}.

Strings can become superconducting, with bosonic or fermionic charge carriers flowing along the strings \cite{Witten:1984eb}. 
Fermionic superconducting strings emerge by introducing fermions, $\psi$, coupled to the complex scalar field $\Phi$.   
The fermion zero mode, $\psi^{(0)}$, is bounded to the string \cite{Kiskis:1977vh, Ansourian:1977qe, Nielsen:1976hs, Jackiw:1981ee}, 
as required by the index theorem \cite{Weinberg:1981eu}. 
When an electric field ${\bf E}$ is applied along the string, a current is generated, 
with its divergence proportional to the electric field, $\partial_\mu j^\mu \propto |{\bf E}|$. Remarkably, this current persists even after the electric field is removed, making the strings superconducting. 
The current generation mechanism is described by the phenomenon of anomaly inflow \cite{Callan:1984sa,Bardeen:1984pm,Harvey:2000yg}.

Superconducting strings present a rich and intriguing phenomenon. Their interaction with the 
background electromagnetic field has been extensively studied in the late 1980s \cite{Naculich:1987ci,Kaplan:1987kh,Manohar:1988gv}. 
Recent work has investigated the cosmological history and potential observational signatures of superconducting strings 
\cite{Fukuda:2020kym,Abe:2020ure,Agrawal:2020euj,Ibe:2021ctf}.
In the early universe, electric fields can charge the strings with fermion zero-mode currents. 
The energy of these fermion zero-modes can counterbalance the tension of a string loop, forming a stable configuration  
known as a {\it vorton} \cite{Davis:1988jq,Davis:1988ij,Carter:1993wu,Brandenberger:1996zp, Martins:1998gb,Martins:1998th,Carter:1999an}.

The stability of vortons remains an open question, influenced  by string dynamics (including string loop contraction and radiation),
fermion zero-mode decay, quantum-tunneling, and scattering processes.  
This paper investigates the stability of vortons at both classical and quantum levels. A summary of our findings and previous results 
from the literature is as follows:

\begin{itemize}[leftmargin=*]

\item {\it Classical Stability}: The vorton configuration arises from a balance between the string tension and the zero-mode energy, neglecting the kinetic energy. Using lattice simulations of cosmic string loops, we show that strings can relax into stable vortons by losing their kinetic energy via radiating Nambu-Goldstone bosons associated with the global $\rm U(1)$ symmetry breaking. We also analyze the classical dynamics of fermion zero modes inside a vorton and find that the maximal possible energy of the zero modes is
$m_\psi \sqrt{m_\psi R}$, where $m_\psi$ is the mass of the fermion outside the vorton and $R$ is the radius of the vorton.
For $R\gg m_{\psi}^{-1}$, the maximal energy is much above the Fermi energy inside the vorton and the vorton is classically stable.

\item 
{\it Quantum tunneling}: The conversion of fermion zero modes to continuum states, $\psi^{(0)}\rightarrow\psi$, 
can occur when the energy exceeds the fermion mass, $E>m_\psi$. 
There is disagreement on the critical tunneling 
energy at which the process becomes efficient. \cite{Witten:1984eb} argued that for $E \gtrsim m_\psi$, the fermions are
energetically favored to escape the string, while \cite{Barr:1987ij} (see also \cite{Vilenkin:2000jqa}) proposed a 
critical energy of $m_\psi^2 R $ based on the dispersion relation of the zero modes.
However, our analysis suggests a different conclusion from both results.
By directly computing the quantum tunneling rate using the S-matrix, we show that
the critical energy for efficient zero-mode quantum tunneling is $E\sim m_\psi \sqrt{m_\psi R}$, which aligns with our classical arguments.

\item {\it Vorton decay}:
The decays of fermion zero modes are induced by their coupling to other particles, such as the quarks $q$ and Higgs $h$.
The existing literature presents different results regarding the decay rate of vortons 
\cite{Fukuda:2020kym,Ibe:2021ctf}. \cite{Fukuda:2020kym} considers the process $\psi^{(0)}\rightarrow q+h$, whereas \cite{Ibe:2021ctf} 
analyzes the process $\psi^{(0)}+\delta\Phi\rightarrow\tilde{\psi}\rightarrow q+h$, where $\delta\Phi$ represents string modulations relative to 
a straight string segment, and $\tilde{\psi}$ denotes an intermediate state of $\psi$.
We demonstrate that
the first process occurs generically even for nearly static string configuration while the latter process occurs when the string configuration changes 
non-adiabatically.  

\item{\it Surrounding charged particle scattering}: The electromagnetic field produced by the fermion zero-mode currents along the string can pair-produce charged particles~\cite{Schwinger:1951nm} and
accelerate them. 
These charged particles can scatter the fermion zero mode off the string and trigger the instability of votons~\cite{Agrawal:2020euj}.
We find that to stabilize the vortons places upper bound on the mass of $\Phi$, so that the transverse wave function of the zero mode becomes sufficiently wide that the electromagnetic field produced by the zero-mode current is weak enough.

\end{itemize}

The structure of the paper is following. In \cref{sec:ss}, we introduce the fermion zero-mode solutions in a straight string and a circular string loop and its anomalous current along the string, which is proved to be compensated by anomaly inflow, and explain why QCD axion strings are superconducting. \Cref{sec:vorton} discusses a stable string loop configuration, vorton, and its cosmology, especially its charges induced by thermal plasma. \Cref{sec:classical_stability} and \cref{sec:quantum_stability} are dedicated to demonstrating the classical and quantum mechanical stabilities of vorton respectively. Conclusion is given in \cref{sec:conclusion}.

\section{Superconducting strings\label{sec:ss}}

In this section, we review the fermion zero-mode solution in a straight string background and construct an approximate solution for a circular string loop. To further understand the production of the fermion zero mode along the string and the associated current conservation, we review the phenomenon of anomaly inflow. Finally, we study the superconductivity of QCD axion strings.

\subsection{Fermion zero modes \label{sec:zeromode}}

We review the fermion zero-mode solutions propagating along a global cosmic string. We consider a system consisting of 
a complex scalar field $\Phi$, charged under a global $\rm U(1)$ symmetry, and a Dirac fermion that is chirally charged under this symmetry
and coupled to $\Phi$. The Lagrangian density is
\begin{equation}
\label{eq:L_psi}    
   \mathcal{L}= \partial_\mu \Phi^\dagger \partial^\mu \Phi 
      - \frac{\lambda }{4} \left( |\Phi|^2 - \frac{v_a^2}{2} \right)^2 
   + i\bar{\psi}   \gamma^\mu  {\partial_\mu} \psi-\left(y_\psi \Phi \, \bar{\psi}_L\psi_R+{\rm h.c.}\right) \, , 
\end{equation}
where $\psi_L  = \frac{1- \gamma_5}{2} \psi $ and $\psi_R  = \frac{1 + \gamma_5}{2} \psi$ 
denote the left- and right-handed components of the Dirac fermion $\psi$, respectively.
The global $\rm U(1)$ symmetry is spontaneously broken by the vacuum expectation value~(vev) of the scalar field,
 $\langle \Phi \rangle  = \frac{1}{\sqrt{2}} v_a$. A global string is a classical solution for the scalar field $\Phi$ characterized by 
a non-zero winding number $n_w$. 
A straight string profile along the $z$ direction can be parametrized in the cylindrical coordinate $(\rho, \theta,z)$ as 
\begin{equation}
   \Phi (\rho, \theta, z)  = \frac{v_a}{\sqrt{2}} f(\rho) e^{i n_w \theta} \equiv  \phi(\rho ) e^{i n_w \theta} \, , 
\label{eq:Phirho}
\end{equation}
where $f(\rho)$ is the radial profile of the string.  
At large distance, $ \rho \to \infty$, 
the absolute value of the scalar field approaches its vev, meaning $f(\infty ) = 1$, while 
at the origin where $\rho = 0 $, we have $f( 0  ) = 0$. 
The string profile can be approximated analytically as 
$f(\rho) \simeq \tanh (m_\phi \rho)$, where $m_\phi=\sqrt{\frac{\lambda}{2} } v_a$ is the mass of the scalar. For later convenience, we define string core size $\delta=\frac{1}{m_\phi}$.

In the two spatial dimensions perpendicular to the string, the fermion and the vortex form a bound state.
The fermion in the bound state is a zero mode (zero energy) state satisfying 
the following Dirac equation,  
\begin{eqnarray}
  i ( \partial_1 \gamma^1 + \partial_2 \gamma^2) \psi_L^{(0)} &= y_\psi \frac{v_a}{\sqrt{2}} f (\rho )  e^{i \theta} \psi_R^{(0)} \, , 
   \\
   i ( \partial_1 \gamma^1 + \partial_2 \gamma^2) \psi_R^{(0)} &= y_\psi \frac{v_a}{\sqrt{2}} f(\rho )  e^{ - i \theta}  \psi_L^{(0)} \, ,
\end{eqnarray}
where we assume a winding number $n_w = 1$ of the string.
The zero-mode solution can be found by using the relations 
$ i \gamma^1 \gamma^2 \psi_L^{(0)} =  \psi_L^{(0)}$ and $ i \gamma^1 \gamma^2 \psi_R^{(0)} =  -\psi_R^{(0)}$, 
\begin{equation}
   \psi^{(0)} = \alpha( t, z ) \, \eta  e^{-m_\psi\int_0^\rho f(\sigma)d\sigma } \, , 
   \quad   \eta = \frac{1}{\sqrt{2}} ( 1, 0, 0, i )^T \, .
\end{equation}
The fermion zero-mode states propagate along the string. 
Due to the separation of variables, the Dirac equation is simplified as $(\partial_0 - \partial_3) \alpha ( z, t) = 0 $,
indicating that the zero-mode states are left-moving along the z-direction at the speed of light, i.e., $\alpha(t, z) = \alpha( t+z)$.

To describe the anti-particle of the fermion, we apply charge conjugate to the fermion field, $\psi \to - i \gamma^2 \psi^*$. 
In the z-direction, the equation for $\alpha$ remains unchanged, indicating that the anti-fermion is also a left-mover.

After quantization, the zero mode field can be expanded as \cite{Harvey:2000yg},
\begin{equation}
 \psi^{(0)}=  \left[ \int_0^\infty \frac{d p}{2\pi} \, \eta \, \hat{a}_p  
      + \int_{-\infty}^0 \frac{d p}{2\pi} \, \eta \, \hat{b}_{-p}^\dagger \right] e^{ - i p ( t + z ) }  {\cal F} ( \rho) \, , 
\label{eq:psi0_quan}
\end{equation}
where $\hat{a}_p$ annihilates a particle with momentum $p$ along the string, 
and $\hat{b}_p^\dagger$ creates a particle with momentum $p$ in the same direction. 
The two operators satisfy the usual anti-commutation relation.
$\mathcal{F} (\rho) $ is normalized in the transverse plane as
\begin{equation}
\label{eq:zm-transverse-wf}
  \mathcal{F}(\rho)=\mathcal{N}e^{-m_\psi\int_0^\rho f(\sigma)d\sigma },~~  
       2 \pi \int_0^{\infty}  \, |\mathcal{F}(\rho)|^2
       \rho d\rho
       =1 \, ,
\end{equation}
where $\mathcal{N}$ is a normalization constant.

The width of the transverse wavefunction depends on the mass hierarchy between $m_\phi$ and $m_\psi = y_\psi v_a/\sqrt{2}$.
First note that $f(\rho)$ is approximated by $m_\phi \rho$ and $\int_0^\rho f(\sigma) d \sigma \simeq m_\phi \rho^2/2 $ for $\rho \ll m_\phi^{-1}$.
When $m_\phi \ll m_\psi$, $\mathcal{F}(\rho)$ is exponentially suppressed for $\rho >(m_\phi m_\psi)^{-1/2}$. When $m_\phi \gg m_\psi$, $(m_\phi m_\psi)^{-1/2} > m_\phi^{-1}$, so the approximation on $f(\rho)$ breaks down. We can instead use the approximation $f(\rho) \simeq 1 $ for $\rho \gg m_\phi^{-1}$.  The exponential suppression of $\mathcal{F}(\rho)$ begins at $\rho\sim m_{\psi}^{-1}$. To sum up, the spreading of the zero mode in the transverse direction is
\begin{equation}
\label{eq:zm_spreading}
   \delta_\psi= \begin{cases}
        \frac{1}{\sqrt{m_\phi m_\psi}} & : m_\phi \ll m_\psi \\
        \frac{1}{m_{\psi}} & : m_\phi \gg m_\psi. 
    \end{cases}
\end{equation}

\subsection{Zero modes in a circular string loop\label{sec:circular}}

\begin{figure}[tbp]
\centering  
    \begin{subfigure}[b]{0.7\textwidth}
        \centering
        \includegraphics[width=1.1\textwidth]{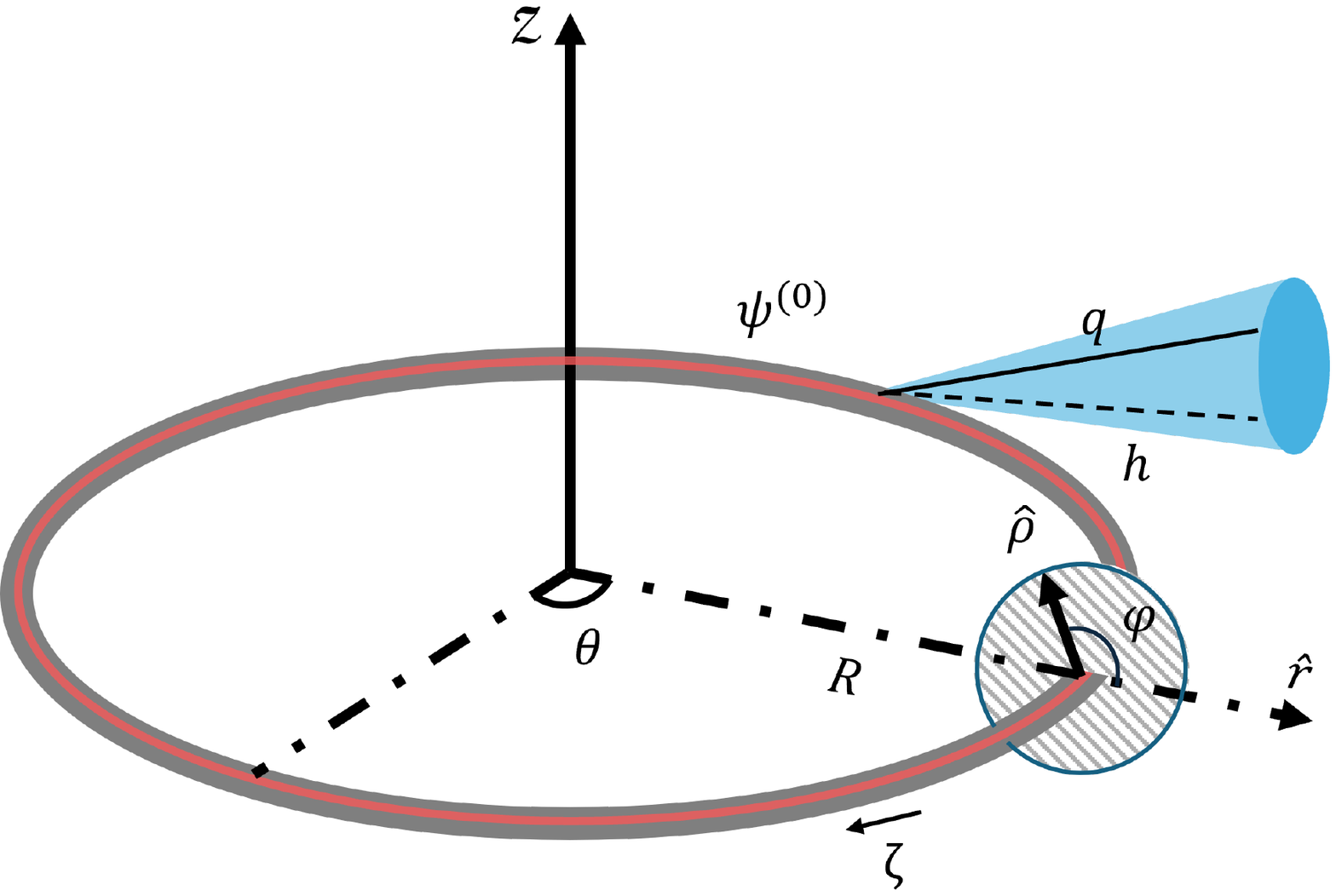}
    \end{subfigure}
    \caption{\label{fig:coord-loop} The coordinate setup of a circular string loop. The string loop is denoted by a gray line, while the red line represents zero modes. The phase space of the decay of $\psi^{(0)}\rightarrow q+h$ is confined in a cone (blue). }
\end{figure}

We extend the analysis on a string string to a circular string loop, relevant for studying the vorton stability. 
The circular string loop with radius $R$ lies in the $x-y$ plane, centered at the origin. 
For this geometry, we employ cylindrical coordinates $(r, \theta , z )$ and define local transverse coordinates along the string as $(\rho, \varphi)$, as illustrated
in \cref{fig:coord-loop}.
Following the Abrikosov ansatz \cite{Abrikosov:1956sx} (detailed in \cref{sec:aa}), the string profile is approximated in the same form as 
the straight string, 
with the correction of $\mathcal{O}(\delta/R)$, $\Phi(r,\theta,z)\simeq \frac{v_a}{\sqrt{2}}f(\rho)e^{-i\varphi}$, 
where $\rho=\sqrt{(r-R)^2+z^2}$ and $\varphi=\tan^{-1}(z/\rho)$ are the local radial distance from the string core and the local azimuthal angle, 
respectively. Note that we choose the winding number of the string $n_w=1$ because this corresponds to a left-moving zero-mode solution in a circular string loop, which propagates along the $-\hat{\theta}$ direction.%
\footnote{The local azimuthal angle $\varphi$ defined for a circular string loop, shown in in the \cref{fig:coord-loop}, corresponds to $-\theta$ in a straight string set up discussed in \cref{sec:zeromode}.}

Similarly, the Dirac equation for the zero mode in the transverse plane of the string is given by,%
\footnote{
To be precise, the masses of fermions trapped inside loop configurations are non-zero. By ``zero mode", we mean the mode whose mass approaches zero as the radius of the loop is taken to be infinity. See~\cite{Chu:2007xh} for the argument supporting the existence of such zero-modes.}
\begin{equation}
i(\gamma^r\partial_r+\gamma^z\partial_z)\psi_L^{(0)}=y_\psi\frac{v_a}{\sqrt{2}} f(\rho ) e^{-i\varphi}\psi^{(0)}_R,
\end{equation}
\begin{equation}
i(\gamma^r\partial_r+\gamma^z\partial_z)\psi_R^{(0)}=y_\psi\frac{v_a}{\sqrt{2}} f(\rho) e^{i\varphi}\psi^{(0)}_L \, .
\end{equation}
Here, the $\gamma$-matrices in cylindrical coordinates are given by
$\gamma^r = \cos \theta \gamma^1+\sin\theta \gamma^2$, $\gamma^\theta=-\sin\theta \gamma^1+\cos\theta \gamma^2$,
and $\gamma^z = \gamma^3$. 
Using the ansatz,
$ i \gamma^r \gamma^z \psi_L^{(0)} =  -\psi_L^{(0)}$, and $ i \gamma^r \gamma^z \psi_R^{(0)} =  \psi_R^{(0)}$, 
we solve the Dirac equations to obtain,
\begin{equation}
\label{eq:zm_solution_circular}
    \psi^{(0)}=\alpha(t+R\theta)\eta(\theta)e^{-m_\psi\int_0^{\rho}f(\sigma)d\sigma},~\eta=\frac{1}{2}(-ie^{-i\theta/2},e^{i\theta/2},ie^{-i\theta/2},e^{i\theta/2})^T \, .
\end{equation}
In the longitudinal direction, the Dirac equation becomes 
\begin{equation}
\label{eq:DiraceqL}
    \gamma^0\eta \, \partial_t\alpha+\frac{1}{r}\gamma^\theta \,  \partial_\theta(\eta\alpha)=0.
\end{equation}
Neglecting the $\theta$-dependence of $\eta$, this reduces to $(\partial_t-\frac{1}{r}\partial_\theta)\alpha=0$, which implies
$\alpha=\alpha(t+R\theta)+\delta\alpha$. 
The term $\delta\alpha$ accounts for the mismatch in the exact solution for $|r - R| \leq \delta$. 
Considering the $\theta$-dependence of $\eta$, the partial derivative of the second term in \cref{eq:DiraceqL} introduces an extra 
$\frac{\alpha}{r}\gamma^\theta\partial_\theta\eta\rightarrow\pm\frac{i\alpha}{2r}$. 
However, the amplitude of this extra term, which corresponds to the contribution of the fermion spin to the angular momentum, is negligible compared to 
$\frac{1}{r}\partial_\theta\alpha \sim \frac{1}{r}\partial_\theta e^{ ip(t+R\theta)}=\frac{i}{r}(p R)\alpha$,
under the condition $n\equiv p R\gg 1$, where $n\in\mathbb{Z}$ is the quantized energy level of zero modes.
The condition is always satisfied when considering the vorton dynamics. 
After quantization, the zero mode decomposes into creating and annihilating parts as
\begin{equation}
 \psi^{(0)}=  \left[ \int_0^\infty \frac{d p}{2\pi} \, \eta \, \hat{a}_p  
      + \int_{-\infty}^0 \frac{d p}{2\pi} \, \eta \, \hat{b}_{-p}^\dagger \right] e^{ - i p ( t + R \theta  ) }   {\cal F} ( \rho) \, ,
\label{eq:psi0_quan}
\end{equation}
where ${\cal F} ( \rho)$ takes the same form as in \cref{eq:zm-transverse-wf}, consistent with the Abrikosov ansatz.

When the energy of the zero mode $ E\simeq p$ is larger than the mass of the fermion outside the string $m_\psi= y_\psi v_a/\sqrt{2}$, the zero modes on the string loop are actually not a stable energy eigenstate. Indeed, the Dirac equation of the fermion far outside the string loop is that of a fermion with a constant mass $m_\psi$. The wavefunction satisfies the wave equation $(\partial^2- m_\psi^2)\psi =0$, which is $\nabla^2 \psi = (-E^2 + m_\psi^2)\psi$. For $-E^2 + m_\psi^2<0$, the wave equation does not have a solution that decays rapidly at spacial infinity, which means that a stable bound state with $E> m_\psi$ does not exist.%
\footnote{
For straight strings along the $z$ direction, the wave equation far outside the string is satisfied by $\psi\sim e^{-iE(z+t)}$, and the fermion does not escape from the string.  
}
The situation is similar to that of typical tunneling problems in quantum mechanics, where a particle in a local minimum is not absolutely stable and eventually tunnels into the absolute minimum. We compute the decay rate of the approximate zero mode in Section~\ref{sec:qt}.

\subsection{Fermion zero modes and anomaly inflow \label{sec:anomaly}}

In this section, we examine the production of fermion zero modes along the string, explain why the string is superconducting, and 
establish total current conservation from a $3+1$D perspective.
The system can be viewed as a $1+1$D field theory on the string, coupled to a $3+1$D field theory in the bulk.
We consider only left-moving electromagnetic charged chiral fermions on the string, and the results can be easily generalized to include color-charged fermions.
Note that this setup differs from the superconducting string containing equal numbers of left- and right-movers, 
discussed in Witten's seminal paper \cite{Witten:1984eb}.

The fermion zero modes are effectively described by a fermion in $1+1$D coupled to a ${\rm U}(1)$ gauge field. 
The consistent anomaly of the current conservation in $1+1$D is~\cite{Alvarez-Gaume:1983ict,Naculich:1987ci}
\begin{equation}
   \partial_\mu j_{1+1}^\mu =  - \frac{e^2} {8 \pi } \epsilon^{ab} F_{ab} \, \delta^{(2)}(x_\perp) \, , 
\end{equation}
where the spacetime index  $a = (0, 3)$, $e$ is the gauge coupling constant, and $ \delta^{(2)}(x_\perp)$ reflects the localization of the current to the string. 
Applying an electric field along the string generates a fermionic current.
Even when the electric field is switched off, the current on the string is conserved. Therefore, the string is superconducting.

From the $3+1$D perspective, the non-conservation of the $1+1$D current arises due to anomaly inflow from the bulk. 
This anomaly inflow, far from the string core, can be calculated using the method of Goldstone and Wilczek \cite{Goldstone:1981kk} 
or axion electrodynamics \cite{Sikivie:1984yz,Wilczek:1987mv}\,\footnote{\Cref{app:axion-QED} gives an example of using axion-QED to understand the anomaly inflow around a straight axion string.}. 
The bulk current off the string is given by $j^\mu = \frac{e^2}{ 16\pi^2} \epsilon^{\mu \nu \lambda \sigma} 
\partial_\nu \theta F_{\lambda \sigma}$. 
A divergence of the current matches twice the fermionic current on the string, $\partial_\mu j^\mu = - 2 \partial_\mu j_{1+1}^\mu $,
so that the total current is still not conserved. 
To reconcile this discrepancy, which originates from sharply separating the bulk from the string, a ``bump function" $g(\rho)$ in the effective action is introduced \cite{Harvey:2000yg},
\begin{equation}
   S_{\rm eff}  = \frac{e^2}{ 16\pi^2}  \int d^4 x \left( 1+ g (\rho) \right)  \epsilon^{\mu \nu \lambda \sigma} 
         \partial_\mu \theta  A_\nu F_{\lambda \sigma} \, .
\end{equation}
The gauge invariance relates the bump function to the fermion zero mode profile, ${\cal F}(\rho)$, defined in \cref{eq:zm-transverse-wf},
\begin{equation}
   g (\rho ) =  - 1 + 2 \pi  \int_0^\rho {\cal F}^2 ( \rho ) \rho d \rho \, ,
\end{equation}
leading to the boundary conditions $g ( 0 ) = - 1$, $g ( \rho \to \infty ) = 0 $, and $g'( 0 ) = 0 $. 
The currents derived from this effective action
is decomposed into the Goldstone-Wilczek current (or Hall current), localized off the string core, 
\begin{equation}
   j^\mu_{\rm GW} = \frac{e^2}{ 8\pi^2} ( 1+ g (\rho) ) \epsilon^{\mu \nu \lambda \sigma} \partial_\nu \theta F_{\lambda \sigma}  \, .
\end{equation}
and an effective current localized on the string,
\begin{equation}
      j^\mu_A = \frac{e^2}{ 8\pi^2} \epsilon^{\mu \nu \lambda \sigma} \partial_\nu \theta \partial_\lambda g ( \rho )  A_{\sigma}  
               =  - \frac{e^2}{ 4\pi} \epsilon^{ \mu b} A_b   \delta^{(2)} ( x_\perp ) \, ,
\end{equation}
where $\mu =0$ or $3$.
The total current, including the contributions from $ j^\mu_{\rm GW} $, $j^\mu_{A}$, and $j^\mu_{1+1 }$, is conserved, 
\begin{equation}
   \partial_\mu j^\mu_{\rm GW}  + \partial_\mu j^\mu_{A}  + \partial_\mu j^\mu_{1+1 }   = 0 \, .
\end{equation}
The effective current on the string is $j^\mu_{\rm string} = j^\mu_{A} + j^\mu_{1+1}$.
When the background electric field is turned off, the Goldstone-Wilczek current vanishes.

\subsection{QCD axion strings are superconducting\label{sec:QCDsting_sc}}

Superconductivity emerges as a generic feature of comic strings, 
and we emphasize that the QCD axion strings are superconducting. Readers interested in QCD axion physics can view the rest of the paper as a detailed study of superconducting QCD axion strings. 
Those focused on generic string physics can consider it as an exploration of superconducting strings more broadly.

We focus on a class of QCD axion models, KSVZ-type QCD axion \cite{Kim:1979if,Shifman:1979if}. 
Cosmic strings arise from the spontaneous breaking of the global Peccei-Quinn (PQ) symmetry, $\rm U(1)_{PQ}$. 
These strings support fermion zero modes due to the KSVZ fermion, $\psi$, charged under both $\rm U(1)_{PQ}$ and the QCD color group, 
$\rm SU(3)_c$.
We further assume a post-inflationary scenario, where the PQ symmetry is spontaneously broken after inflation.
This is achieved by a sufficiently high temperature of the universe such that the thermal correction to the potential of the PQ breaking field dominates over the vacuum potential, or the Hubble scale during inflation above the PQ breaking scale.
Under these conditions, cosmic strings form during the $\rm U(1)_{PQ}$ phase transition after inflation. The KSVZ fermion $\psi$ is massless before the phase transition and can be abundantly produced from the thermal bath or by quantum fluctuation during inflation.

In KSVZ-type models, the presence of $\rm SU(3)_c$-charged fermions $\psi$ introduces a potential issue of fractionally charged relics \cite{DiLuzio:2016sbl,DiLuzio:2017pfr}. 
To circumvent this, the fermion should belong to a specific representation under the Standard Model gauge group.
For instance, considering the fermion in the fundamental representation of $\rm SU(3)_c$ and a singlet under $\rm SU(2)_L$, 
$-\frac{1}{3}+\mathbb{Z}$ hypercharge is required to address this issue.  
By choosing the integer $Z= 0$, the fermion can couple to the Standard Model Higgs $h$ and quarks $q$ via Yukawa couplings, 
\begin{equation}
\label{eq:L_D}
    \mathcal{L}=- y_D \bar{q}_L h \psi_R + {\rm h.c.} 
\end{equation} 
For phenomenological viability, if $\psi$ is abundantly produced in the early universe, it should decay before Big Bang Nucleosynthesis~(BBN), which places a lower bound on the Yukawa coupling, 
$y_D \gtrsim 5\times 10^{-16}(10^{10}{\rm GeV}/m_\psi)^{1/2}$. 
This interaction also provides a decay channel for the fermion zero modes. As we show in \cref{sec:quantum_stability}, the resultant decay rate is too large for vortons to be stable in the cosmological time scale.
The vorton can be stable if $q$ or $h$ in \cref{eq:L_D} is a massive beyond-Standard Model particle, for which we find exponential suppression of the decay rate. $q$ and $h$ in the following discussion should be understood as  generic fermion and scalar particles.

As a result, the KSVZ fermion zero modes on the axion strings carry both color and $\rm U(1)_{Y}$ charge. We focus on $\rm U(1)_{em}$ rather than $\rm U(1)_Y$ as we consider the scale below the electroweak scale. 
The QCD axion strings are superconducting, which arises from either the color or electromagnetic charge. 
It is revealed by the ${\rm U}(1)_{\rm PQ}{\rm U}(1)_{\rm em}^2$ and ${\rm U}(1)_{\rm PQ}{\rm SU}(3)_{\rm c}^2$ anomalies. 
The anomaly term is given in the effective Lagrangian by integrating out the fermions,
\begin{equation}
\label{eq:L_axion}
    \mathcal{L}=     
      \frac{E_{\mathcal{A}} }{ N}\frac{ e^2 }{32 \pi^2}\frac{a}{v_a} F_{\mu\nu}\tilde{F}^{\mu\nu}
         +   \frac{ g_s^2 }{32 \pi^2}\frac{a}{v_a} G^a_{\mu\nu}\tilde{G}^{a\mu\nu},
\end{equation}
where $g_s$ is the strong coupling constant, 
$\tilde{F}_{\mu\nu}=\frac{1}{2}\epsilon_{\mu\nu\rho\sigma}F^{\rho\sigma}$ and 
$\tilde{G}^a_{\mu\nu}=\frac{1}{2}\epsilon_{\mu\nu\rho\sigma}G^{a\rho\sigma}$.
Here, the anomaly coefficient $E_{\mathcal{A}}=\sum_{f_R} Q_{{\rm PQ},f_R}Q_{f_R}^2- \sum_{f_L} Q_{{\rm PQ},f_L}Q_{f_L}^2$, with $Q_{{\rm PQ},f}$ denoting the PQ charge and 
$Q_f$ is the electric charge of the heavy fermion. $f_L$ and $f_R$ and left-handed and right-handed fermions, respectively. The sum runs over flavor and color indices.
For a single fermion, $Q=-\frac{1}{3}$, $Q_{{\rm PQ}}=\pm\frac{1}{2}$ (suppose the PQ scalar field carries a unit PQ charge, and thus left-handed and right-handed fermions carry chiral PQ charges), we find $E_{\mathcal{A}}=\frac{1}{3}$ and $N =\frac{1}{2}$, which gives the domain wall number 1 to avoid the domain wall problem in the universe \cite{Sikivie:1982qv}.

When an electric field is applied along the string, fermion zero modes in color-singlet states are generated. 
Similarly, applying a color electric field along the string generates zero modes forming $\rm U(1)_{em}$-singlet. 
These phenomena highlight the superconducting nature of QCD axion strings.

\section{Vorton\label{sec:vorton}}

In this section, we introduce the stable solution of superconducting strings, known as a {\it vorton}.  
We analyze the critical length of superconducting strings (the size of vorton) and discuss the process that charges the strings in the universe.

\subsection{Critical length}
Suppose a superconducting string carries a total charge $Q$, uniformly distributed along its length $L$, with the 
charge density per unit length given by, 
\begin{equation}
\lambda_Q = Q / L \, . 
\label{eq:rhopsi}
\end{equation}
Considering the one-dimensional density of states, the charge density can also be expressed as
\begin{equation} 
   \lambda_Q = e \int_0^{p_F} \frac{ d p }{ 2 \pi } = \frac{e \, p_F }{ 2 \pi}    \, ,
\label{eq:rhopsi2}
\end{equation} 
where $p_F$ is the Fermi momentum. Since the zero mode travels at the speed of light along the string, 
the Fermi momentum equals Fermi energy $p_F = \epsilon_F$.
For a zero mode originating from the KSVZ fermion with an electromagnetic charge $q_\psi$ , 
the density is multiplied by a factor of $|N_c q_\psi|$.
Combining \cref{eq:rhopsi,eq:rhopsi2}, we find that the Fermi energy scales inversely with the string length $L$ for a fixed total charge $Q$, 
\begin{equation}
\label{eq:Fermienergy}
   \epsilon_F = \frac{ 2\pi Q} {e L} \, . 
\end{equation}
The zero-mode energy density per unit length is given by 
\begin{equation}
   \varepsilon_{\rm zero-mode} = \int_0^{\epsilon_F} \frac{ p \, {\rm d} p } {  2 \pi }  = \frac{\epsilon_F^2} { 4 \pi } \, \label{eq:zeromodeenergy}.
\end{equation}
and the total energy from the zero mode becomes
\begin{equation}
   V_{\rm zero-mode} =  \varepsilon_{\rm zero-mode} \times L = \frac{\epsilon_F^2} { 4 \pi } L  = \frac{ \pi Q^2 }{ e^2 L } \, .
\end{equation}
This energy decreases with increasing length $L$.
In contrast, the string mass is approximately proportional to the string length,
\begin{equation}
    M_{\rm string}=\mu L= \pi v_a^2 L \ln\left(\frac{L}{\delta}\right) \, , 
\end{equation}
where $\mu$ is the string tension and $\delta$ represents the core thickness of the string.
For a stationary string, the total energy is given by the sum of the string mass and the zero-mode energy,
\begin{equation}
    V_{\rm tot}=  M_{\rm string} + V_{\rm zero-mode} \, . 
\end{equation}
Here we do not include the Coulomb potential from the electric field around string \cite{Agrawal:2020euj}, which is subdominant compared with the Fermi energy with a moderate size
of $\ln{(L/\delta_\psi)}$.

There exists a critical length $L_c$ at which the forces from the two potentials balance, leading to a stable configuration, known as a {\it vorton}.
Neglecting the logarithmic term in the string tension, the size of the vorton is approximately given by 
\begin{equation} 
\label{eq:vorton_Lc}
    L_c \sim \frac{ Q} { e v_a}    
\end{equation} 
At this critical length, the Fermi energy is
\begin{equation} 
\epsilon_F \sim 2 \pi v_a \,\label{eq:criticalfermienergy} .
\end{equation}
The Fermi energy is typically larger than the fermion mass $m_\psi$, which is at the most $4\pi v_a$ even in strongly coupled models and is much smaller in perturbative models.

One may wonder if the fermion with an energy larger than $m_\psi$ may rapidly jump out of the vorton. In \cref{sec:qt}, we argue that the rate of such a process is exponentially suppressed.

\subsection{Charging the strings}
The strings form after the global $\rm U(1)$ symmetry breaking in the early universe. Interactions between strings and their evolution 
result in a few long strings per Hubble volume, entering the scaling regime. We consider the charging of superconducting strings 
in the Standard Model plasma.

An electric field induces a charge on the string, with the charging rate per unit length per time given by 
\begin{equation}
   \frac{ {\rm d} Q } {{\rm d} t {\rm d } \ell }  = \frac{ e^2 |\bold{E}_l| } { 2 \pi }\, . 
\end{equation}
The plasma-induced electric field has a fixed direction over one coherent time, $t_{\rm coh}\sim T^{-1}$, and 
within a coherent length, $\ell_{\rm coh}\sim T^{-1}$, where $T$ is the plasma temperature.%
\footnote{Non-thermal primordial magnetic fields, if produced by some mechanisms, can charge the string as well which is carefully studied in \cite{Agrawal:2020euj}, but we do not consider this possibility in this paper.}
The typical amplitude of the longitudinal electric field is estimated by $|\bold{E}_l|\sim T^2$.

\begin{figure}
\centering  
  \includegraphics[width=0.8\textwidth]{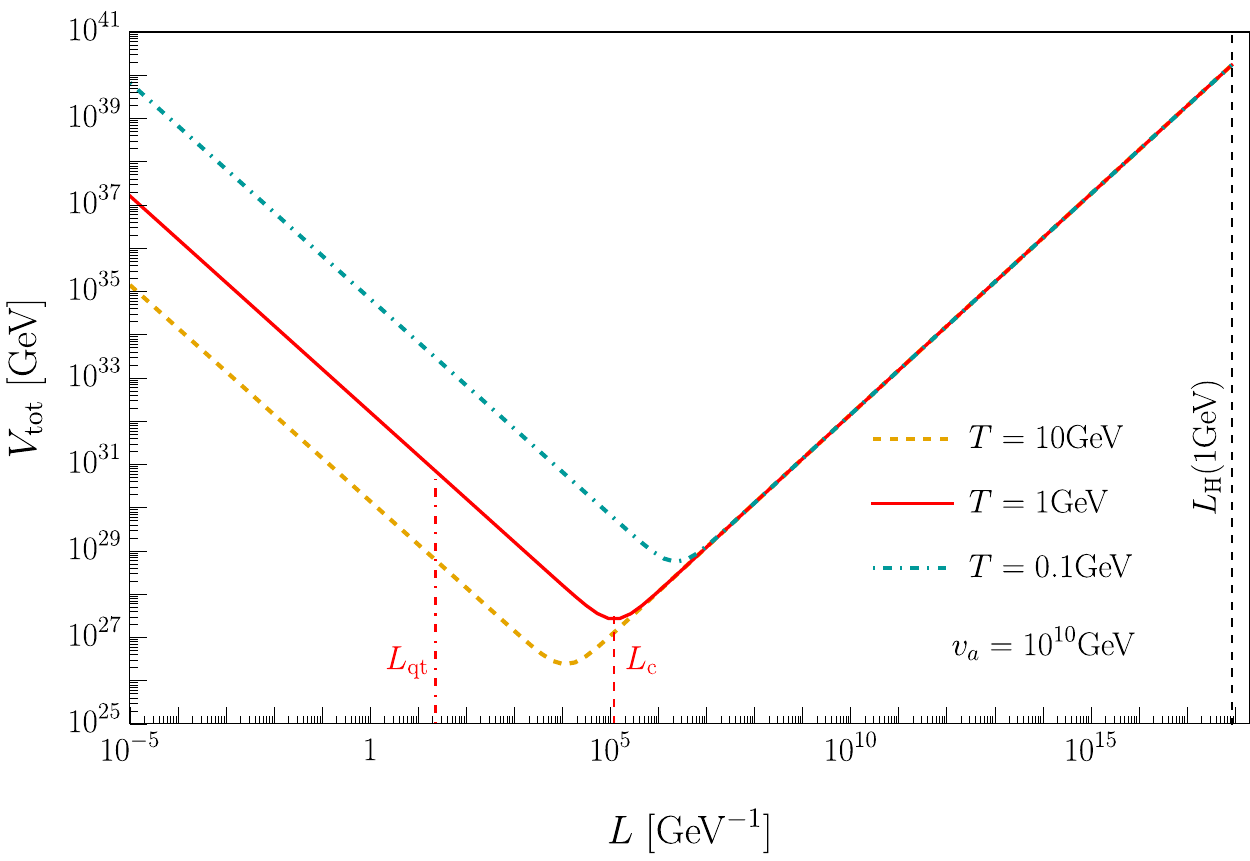}
 \caption{\label{fig:Etot} The total energy as a function of the loop size of a vorton. We choose $v_{a}=10^{10}$\,GeV, Yukawa coupling $y_\psi=1$, and three different temperatures. We assume that the initial size of a stationary string loop is Hubble size, and there is no charge leakage and other dissipation processes during the shrinkage of the string loop. The critical length ($L_{\rm c}$) of $T=1$\,GeV case is shown by a red dashed line. The loop size where the Fermi energy is equal to the maximal zero-mode energy is shown by a red dot-dashed line $(L_{\rm qt}$) for $T=1$\,GeV case. To the left region of $L_{\rm qt}$, the string loop is unstable because the zero-mode energy exceeds the maximal energy and fermion zero modes escape from the string.
 The Hubble size at $T=1$\,GeV is shown by a black dashed line.}
\end{figure}

For a long string with a Hubble size over a Hubble time at the temperature $T$, 
the total charge induced on the string loop can be estimated by a root-mean-square calculation. 
This accounts for a ``random walk" process, where
the step size corresponds to the coherent length and time, 
\begin{equation}
\label{eq:Qrms}
   Q_{\rm r.m.s}  \simeq 
      \sqrt{ \frac{ H^{-1} } { \ell_{\rm coh}} }
      \sqrt{ \frac{ H^{-1} } { t_{\rm coh}} } 
      \frac{ e^2 T^2}{ 2 \pi }  \sim \frac{e^2}{ 2 \pi }  \frac{M_{\rm pl}}{T} 
      \simeq 10^{15} e \left( \frac{ {\rm GeV} } { T} \right)  \, ,
\end{equation}
where $M_{\rm pl}$ is the Plank mass. The length of the vorton formed with this charge is estimated as 
\begin{equation}
\label{eq:Lc}
  L_c \simeq  \frac{  Q }{ e v_a } \simeq   
      10^{5}  \left( \frac{ {\rm GeV} } { T} \right)  \left( \frac{ 10^{10} {\rm GeV} } { v_a} \right)  {\rm GeV}^{-1}\, .
\end{equation}
At a temperature of approximately $1\,{\rm GeV}$, around which the QCD axion strings disappear by the domain wall tension, 
a long string obtains a total charge of about $10^{15} e$ in one Hubble time. 
If the long string relaxes into a single vorton, the charge remains of the same order,
as charge leakage or dissipation processes are negligible at this temperature. 
The critical length at this stage is approximately $L_c \sim 10^{5}\, {\rm GeV}^{-1}$ for $v_a = 10^{10} \,{\rm GeV}$.  
If string intersections occur, the total charge of the resulting vortons is reduced accordingly.
\Cref{fig:Etot} illustrates the total energy of vortons as a function of their loop size, corresponding to three different thermal bath temperatures:
$0.1\,{\rm GeV}$,
$1\,{\rm GeV}$,
and $10\,{\rm GeV}$.

\section{Vorton stability at the classical level\label{sec:classical_stability}}

Conventionally, vorton solutions assume that the kinetic energy of axion strings is negligible, resulting in static configurations. 
However, in the early universe, vorton states are produced by the shrinking of large string loops. When the kinetic energy of cosmic string loops 
becomes excessive while shrinking, even after reaching a stable vorton radius, the loop continues to shrink.%
\footnote{
Here we consider a circular loop that can contract indefinitely. Typical string loops produced in the early universe are not completely circular. They will not shrink rapidly but oscillate, radiate Goldstone bosons, and shrink gradually without overshooting.
See~\cite{Saurabh:2020pqe} for the simulation showing the gradual shrinkage.
We thus expect that overshooting does not occur for typical string loops even for local strings.
Our analysis of the circular loop below can be considered as the confirmation of the absence of overshooting in global strings even for the extreme configuration that may most easily overshoot. Also, for sufficiently non-circular initial configurations, loops may self-intersect to fragment into smaller loops. That only changes the correspondence between the temperature and the vorton charge discussed in \cref{sec:vorton} by ${\mathcal O}(1)$ factors.   
}

The excessive shrinkage, or overshoot, can trigger vorton instability, as the Fermi momentum of zero modes may exceed the critical value, 
making it energetically favorable for the zero modes to escape the string via quantum tunneling or even classically.

The overshoot, however, may be prevented by the Goldstone boson radiation from the string loops that dissipate the kinetic energy of the string loops. 
We perform simulations of long string loops shrinking to smaller sizes, measuring their kinetic energy relative to the string radius. 
The effect of Goldstone boson radiations is captured in these string simulations.
Our analysis confirms that cosmic strings relax to the critical length, and a stable vorton solution arises through a dynamical process.

Furthermore, we calculate the critical energy of fermion zero modes required to jump out of string classically, given by 
$\epsilon_c \sim m_\psi \sqrt{R m_\psi}$, 
where $R$ is the vorton radius. This classical result is consistent with quantum tunneling calculations detailed in \cref{sec:qt}, 
and support the stability of vortons at the critical length $L_c$.

\subsection{String simulations}

We perform lattice simulations to investigate string dynamics, incorporating circular string loops of various sizes.
The primary goal of these simulations is to quantify the kinetic energy of strings relative to string loop radius.
To simplify the analysis, two approximations are adopted.
First, considering the azimuthal symmetry of circular strings, we reduce 3D lattice simulations to 2D.
Second, we focus on the early stages of string evolution, prior to the dominance of fermion energy,
and therefore exclude fermion zero modes from the string configurations.

The simulations reveal that the global string shrinks with mild relativistic velocities, characterized by an asymptotic Lorentz factor
$\gamma\sim \mathcal{O}(1)$.
As a result, we expect superconducting strings to stabilize at a critical length $L_c $, where
the fermion energy prevents the string from collapsing.

Our initial condition is a static circular string loop in the $x-y$ plane with loop radius $R$ and string core size $\delta$. 
The azimuthal symmetry around the center of the loop reduces the 3D string loop into a 2D vortex-anti-vortex pair. 
In the case of $R\gg\delta$, both vortices can be treated as isolated vortices with opposite winding directions. 
In cylindrical coordinates $(r,\theta, z)$ illustrated in \cref{fig:coord-loop}, the initial vortex-anti-vertex pair can be written as:
\begin{equation}
   \begin{split}
    \Phi_{\text{v}} &= 
    \frac{v_a}{\sqrt{2}} f\big(\sqrt{(r-R)^2+z^2}\big) e^{i\varphi_{\text{v}(r,z)}} \, ,\\ 
    \Phi_{\bar{\text{v}}} &= 
    \frac{v_a}{\sqrt{2}} f\big(\sqrt{(r+R)^2+z^2}\big) e^{i\varphi_{\bar{\text{v}}(r,z)}} \, ,
   \end{split}
\end{equation}
where $\Phi_\text{v}$ is the vortex centered at $(R,0,0)$ with a phase $\phi_{\text{v}}$, 
$\Phi_{\bar{\text{v}}}$ is the anti-vortex at $(R,\pi,0)$ with an opposite winding direction, 
and $f$ is the string profile function in \cref{eq:Phirho} and the glabal $U(1)$ symmetry is restored at $f(r=R,z=0)$. 
Note that the angle $\theta$ doesn't appear in the equation above due to the azimuthal symmetry. 
Since these two vortices are widely separated, we take a straight string approximation for the profile function and further simplify it to 
analytical approximation as the initial profile, 
$f(r) \simeq \tanh(m_\phi r)$, 
and the phases $\varphi_\text{v}=\arctan(\frac{r-R}{z})$, $\varphi_{\bar{\text{v}}}=-\arctan(\frac{r+R}{z})$. 
According to Abrikosov ansatz (detailed in appendix \ref{sec:aa}), the complex scalar field $\Phi$ at $t=0$ can be written explicitly as
\begin{align}
    \Phi (t = 0 ) =& \frac{\Phi_\text{v} \Phi_{\bar{\text{v}}}}{v_a/\sqrt{2}}
      \\
=&\frac{v_a}{\sqrt{2}}\tanh(m_\phi\sqrt{(r-R)^2+z^2})\tanh(m_\phi\sqrt{(r+R)^2+z^2})e^{i (\varphi_\text{v}+\varphi_{\bar{\text{v}}})} \nonumber \, ,
\end{align}
and the initial velocity of the strings is set to zero, 
\begin{align}
 \dot{\Phi} ( t= 0 )=0.
\end{align}

By decomposing the complex field into its real and imaginary components, $\Phi = \frac{\phi_1+ i \phi_2}{\sqrt{2}}$, we solve the equations of motion for 
each component,   
\begin{align}
    \ddot{\phi}_i = \nabla^2 \phi_i - \frac{\lambda}{4}(\sum_{j=1,2} \phi_j \phi_j-v_a^2) \phi_i \, , 
\end{align}
where $i = 1, 2$. 
To evolve the fields on the lattice, 
we use the Crank-Nicholson algorithm \cite{Teukolsky:1999rm} with two iterations, 
and impose absorbing boundary conditions.\footnote{Periodic boundary conditions are unsuitable for this configuration 
because the winding directions break periodicity in vortex-anti-vortex setups.}
The model parameters are set to $\lambda = 2.0$, $v_a=1.0$. 
The lattice grid spans 2048 sites in the $r$-direction and 512 sites in the $z$-direction. 
Simulations are run with a lattice spacing of $\Delta x =\frac{1}{4}$ and a time step of $\Delta t = \frac{1}{16}$.
Decreasing the size of $\Delta x$ and $\Delta t$ further does not change our results. The number of time steps varies with the initial loop size, ranging from 2000 to 4000. We adjust the time scale for each run to ensure the first loop collapse is observed. 

As the string loop evolves, it accelerates and shrinks toward its center. 
We identify the location of the string loop by tracking the lattice site with the minimal value of $|\Phi|$.
The loop radius is determined from the $r$-coordinate of this location. 
Simulations are stopped when the loop size becomes comparable to the core size, $\delta$. 
Across all runs, the string loops achieve an asymptotic velocity with a Lorentz factor $\gamma \sim {\cal O} (1)$
before the radius contracts to $\delta$.
The loss of kinetic energy during this process is attributed to the conversion of it into Goldstone boson radiation energy.

\begin{figure}[htbp]
\centering  
    \begin{subfigure}[b]{0.75\textwidth}
        \centering
        \includegraphics[width=0.75\textwidth]{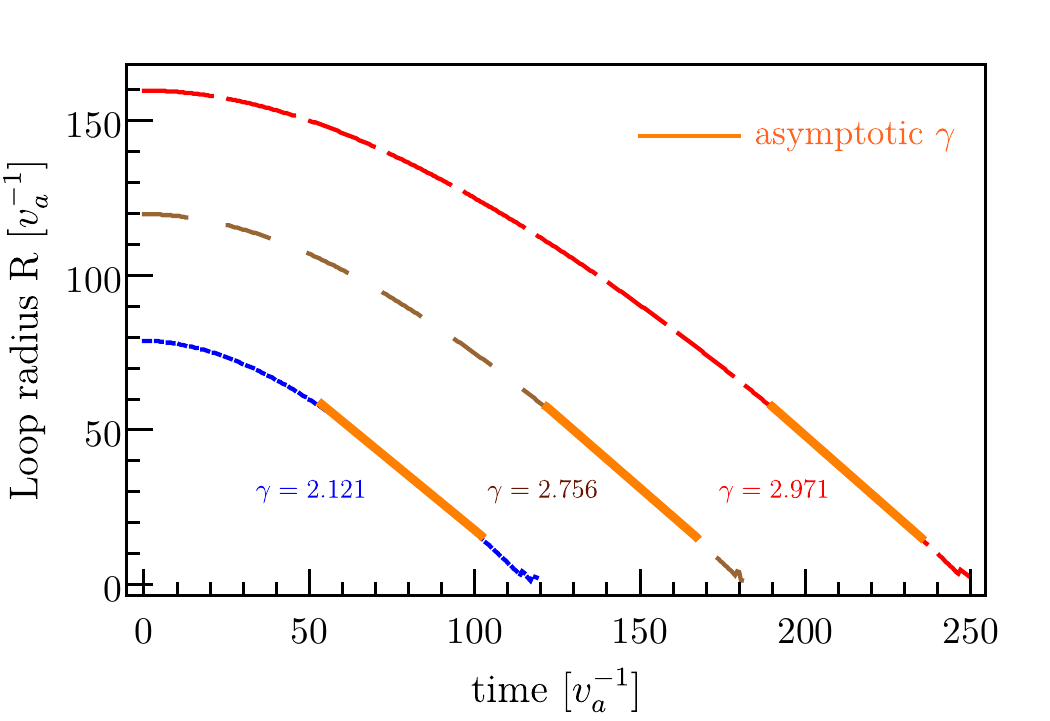}
    \end{subfigure}
    \hfill
  \begin{subfigure}[b]{0.77\textwidth}
        \centering
        \includegraphics[width=0.77\textwidth]{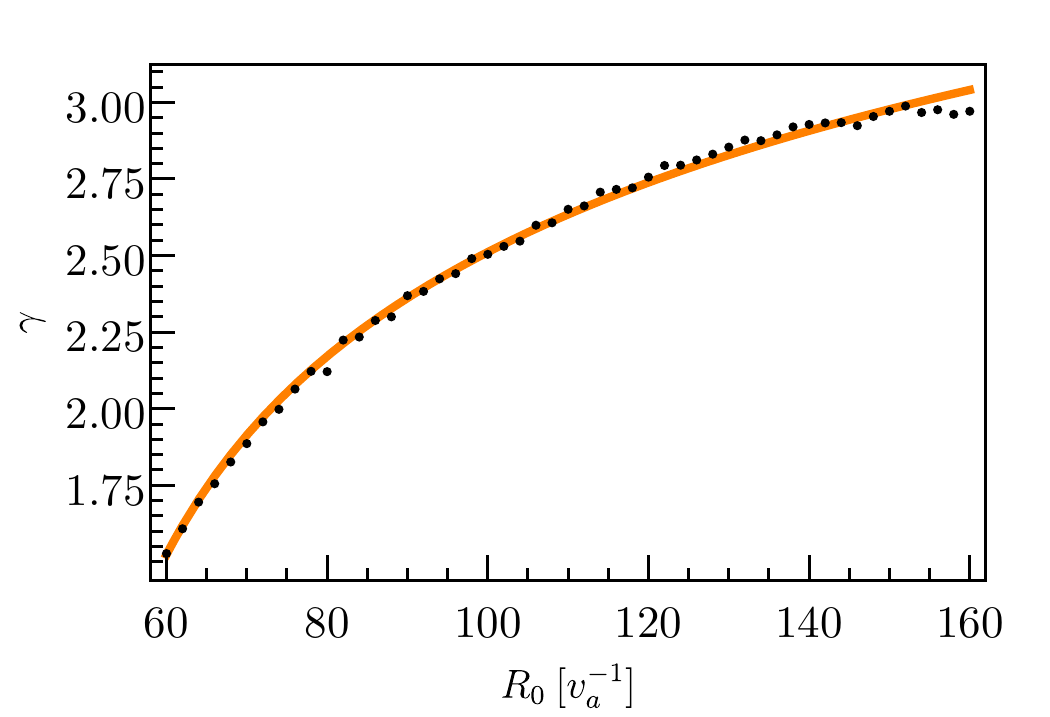}
    \end{subfigure}
\caption{\label{fig:simulation} 
The top panel illustrates three examples of loop evolution, with
radii shrinking from $R_0 = 80, 120, 160$ towards the core size $\delta$. 
The yellow line indicates the asymptotic Lorentz factor achieved by the string loops. 
The bottom panel presents the simulation results, showing the Lorentz factor as a function of 
the initial string loop radius $R_0$. 
}
\end{figure}

Simulation results are presented in \cref{fig:simulation}. 
The top panel shows the evolution of three initial loop radii, $R_0=80, 120, 160$, 
as they shrink toward their centers. In each case, the string achieves an asymptotic velocity before the loop size reaches 
$R \sim \delta$. We use linear regression, represented by the yellow line, to calculate the boost factor $\gamma$ 
as the loop radius decreases from $40 \delta$ to $10\delta$. The bottom panel displays results from 51 simulations with initial loop radii 
$R_0$ uniformly sampled between 60 and 160. These results indicate that the Lorentz factor grows more slowly than 
logarithmically with the loop radius. A best-fit curve is given by $\gamma=10 \tan^{-1}[(\frac{R_0}{\delta}-30)^{0.02}]-8.75$.

While these simulations do not encompass the dynamics of strings shrinking from Hubble scales to the critical length $ L_c $, 
they suggest that the string’s final kinetic energy is comparable to its mass, consistent with an $\mathcal{O}(1)$ boost factor. 
We acknowledge that string evolution is inherently more complex, involving a variety of configurations, string intersections, 
and cosmic string networks. Nonetheless, we argue that these simplified simulations capture the essential physics: 
the backreaction from Goldstone boson radiation plays a critical role in relaxing the strings toward their critical length.

\subsection{Classical stability\label{subsec:overshoot}}

The simulations indicate that the superconducting strings relax to a critical length corresponding to the minimal of the potential 
shown in \cref{fig:Etot}.
The minimal potential results from a balance between the string tension and the energy of the zero modes. 
The string tension can be estimated as $\mu \sim {\cal O}(1)  v_a^2$, while the energy density of the zero modes is related to Fermi energy 
$ \varepsilon \sim  \epsilon_F^2 $ in \cref{eq:zeromodeenergy}. Near the critical length, the Fermi energy 
is given by $\epsilon_F \sim v_a = \sqrt{2} m_\psi / y_\psi$. 
For a Yukawa coupling $y_\psi \sim {\cal O} (1) $, the Fermi energy is approximately equal to the fermion mass $m_\psi$ at $L_c$.

Understanding the stability of the vorton requires determining the critical energy at which zero modes convert into free fermion states 
without suppression. 
Ref.~\cite{Witten:1984eb} argued that at an energy $\epsilon_c \sim m_\psi$, the zero modes 
become energetically favorable to convert to free particle states. It suggests that the stability of vorton at $L_c$ requires a large Yukawa coupling. 
In contrast, Ref.~\cite{Barr:1987ij} used a classical argument to estimate the critical energy as $\epsilon_c \sim m_\psi^2 R$.
In this work, using classical dynamics, we deduce the Fermi critical energy $\epsilon_c$ is $\sim m_\psi \sqrt{R m_\psi}$ 
when the fermion travels along a circular string loop with radius $R$.

The analysis considers the conservation of angular momentum and total energy,
\begin{align}
    R_i p_{\theta,i} &= R_o p_{\theta,o} ,
   \label{eq:Lcons}
      \\
   E^2= p_{\theta,i}^2+p_{T,i}^2 &  =  p_{\theta,o}^2 + p_{T,o}^2 + m_{\psi}^2 ,
   \label{eq:energyconservation}
\end{align}
where
$R$ is the radius of the trajectory of the fermion, and $p_{\theta}$, $p_{T}$ are the longitudinal and transverse momentum.
Subscript $i,o$ denote states inside and outside the string loop, respectively. The conversion process is not in a perfect circular string and its uncertainty is of the order of the spreading of fermion zero mode, $\Delta R = R_o-R_i \sim \frac{1}{m_{\psi}}$.

Using angular momentum conservation in \cref{eq:Lcons}, the difference in radii $R_i$ and $R_o$ leads to a longitudinal momentum difference,
\begin{align}
    p_ {\theta,i} =p_ {\theta,o} \left( 1 + \frac{1 }{m_\psi R} \right). 
\end{align}
If the zero mode with $\epsilon_c$ overcomes the energy barrier in the transverse direction, the transverse momentum outside the vorton should vanish and the excited state only propagates in the tangential direction. Thus the critical energy is acquired by setting $p_{T,o}=0$.
Combining this with energy conservation in \cref{eq:energyconservation}, we find the maximal longitudinal momentum inside the string, 
\begin{align}
    p_{\theta,i} \sim m_\psi \sqrt{m_\psi R},
\end{align}
where we took $p_{T,i} < m_\psi$ to obtain the maximal $p_{\theta,i}$.
So the upper bound for $\epsilon_c$ is
\begin{align}
    \epsilon_c\sim m_\psi \sqrt{m_\psi R} \, .
\end{align}
The critical energy is several orders of magnitude larger than the Fermi energy $\epsilon_F$ in \cref{eq:criticalfermienergy}. The quantum tunneling calculation presented in \cref{sec:qt} 
yields consistent results with this classical-dynamics analysis.

$\epsilon_c \gg \epsilon_F$ guarantees the stability of vorton against possible overshoot.
When the string loop shrinks, because the kinetic energy is only comparable to the mass energy, the shrink will stop when the loop size is smaller than the critical length $L_c$ in \cref{eq:Lc} only by an ${\mathcal O}(1-10)$ factor. The Fermion energy at that loop size is still much smaller than $\epsilon_c$ and hence fermions will not be ejected from the loop. The vorton is eventually stabilized at the critical length $L_c$. In \cref{fig:Etot}, take $T=1$ GeV for an example, we illustrate the overshoot(quantum tunneling) happens when loop size $L<L_{qt}$, and the vorton critical size $L_c$ is four orders of magnitude larger than $L_{qt}$, therefore the initial loop size of Hubble length would shrink into a vorton of size $L_c$ and become stable.

Finally, we comment on the discrepancy between our result and those of~\cite{Barr:1987ij},
which estimated the critical Fermi energy to be several orders of magnitude larger than our result.
This discrepancy arises from their implicit assumption that the momentum is conserved. They computed the energies of the KSVZ fermion inside and outside as a function of their momenta, and obtained the critical energy by equating the two for the same momenta.  
However, the momentum conservation along the string direction is mildly broken by the curvature of the string loop, which makes it easier for the fermion to escape the string.

\section{Superconducting string dynamics: decays and scatterings\label{sec:quantum_stability}}

In this section, we investigate the quantum-level stability of superconducting strings,  
focusing on zero-mode decay, quantum tunneling, and scattering processes.

\subsection{Zero-mode decay\label{sec:zm_decay}}

The decay rate of zero modes is a critical quantity in understanding the dynamics of superconducting string loops.
Zero mode decay is driven by interactions with the scalar field forming the string background and with other particles. 
Once the decay process is initiated, the Fermi pressure within the vorton can no longer counteract the string tension, leading to further 
shrinking of the vorton and dissipation of its energy through Nambu-Goldstone boson radiation. 
The shrinking of the string enhances the decay rate, further accelerating the process.

The decay processes of zero modes into other particles always happen in a curved string background. 
If the time scale of string modulation is much shorter than the relaxation time of the string-zero-mode system, there is another non-adiabatic decay channel for zero modes induced by this sudden string modulation.
To analyze these decay mechanisms, we begin by introducing the string worldsheet formalism, which serves as a foundation for calculating the 
decay rate for a circular string loop (vorton). This formalism aligns with the results obtained from the $3+1$D S-matrix analysis, 
as detailed in \cref{app:s-matrix}, and is extendable to more general string configurations. Using this approach, 
we compute the zero-mode quantum tunneling probability for a one-particle final state, which is discussed in the next subsection.
Finally, we examine the non-adiabatic decay process, identifying its regime
of validity and key features.

\subsubsection{Zero-mode decay: $\psi^{(0)}\rightarrow q+h$ \label{sec:psi2qH}}

We first examine the zero-mode decay process. 
To model the interaction, we employ a Yukawa-like interaction in \cref{eq:L_D} to couple the zero modes with other particles $q$ and $h$, 
which induces the decay process $\psi^{(0)}\rightarrow q+h$.

The decay rate of the zero mode in the presence of a string background can be evaluated using the S-matrix formalism. 
This involves accounting for the finite width of the strings. 
However, under certain conditions, the worldsheet formalism offers a useful simplification. 
This approach approximates the string as a Nambu-Goto string, neglecting its finite width. 
This simplification is valid when the dispersion relations of the intermediate and final states in the decay process  
are independent of the string's internal structure or the zero-mode profile. When the condition is not satisfied, a more careful treatment is
needed. By adopting the worldsheet formalism, 
the $3+1$\,D integral required for the S-matrix calculation reduces to a  $1+1$\,D integral, significantly streamlining the computation.

\paragraph{String worldsheet formalism}

We introduce the worldsheet coordinates $\sigma^a=(\tau,\zeta)$ and express the zero-mode field as 
$\psi^{(0)}(x) \propto \eta(\sigma^a)\mathcal{F}({\bf x_{\perp}})$, 
which generalizes the solutions
described in \cref{sec:zeromode,sec:circular}. 
The interaction between a zero mode that is confined to the string and the bulk particles is effectively localized on the string. 
Consequently, effective coupling $\kappa$ can be defined by integrating over the transverse coordinates,
\begin{equation}
  \kappa= y_D\int d^2{\bf x_{\perp}}\mathcal{F}({\bf x_{\perp}})\sim y_D m_\psi^{-1},
\end{equation}
where $m_\psi^{-1}$ represents the characteristic spreading of the transverse zero-mode wavefunction.
Here we assume $m_\phi > m_\psi$. If $m_\phi \ll m_\psi$, the characteristic spreading becomes $(m_\psi m_\phi)^{-1/2}$ and $\kappa \sim y_D (m_\psi m_\phi)^{-1/2}$.
The Lagrangian for the fermion zero modes can then be written as,
\begin{equation}
    \mathcal{L}_{\rm int}=-\kappa \int d^2\sigma \sqrt{-\gamma}\,\delta^4(x^\mu-x^\mu(\sigma^a))(h (x) \, \bar{\psi}^{(0)}_R(\sigma^a) 
      \, q_L(x) 
      + h.c. ) \, ,
\end{equation}
where $\gamma_{ab}=g_{\mu\nu}x^\mu_{,a}x^\nu_{,b}$ is the induced metric on the string worldsheet and 
$\gamma=\text{det}(\gamma_{ab})$.
For convenience, we adopt the conformal gauge of the 2D worldsheet, where
the worldsheet time coordinate is set to be physical time $\tau=t$, and the following conditions hold, 
\begin{equation}
   \dot{\bf x}\cdot {\bf x}'=0, \quad |\dot{\bf x}|^2+|{\bf x}'|^2= 1  \, . 
\label{eq:conformalG}
\end{equation}
Under these conditions, the determinant simplifies to $\sqrt{-\gamma}=|{\bf x}'(t, \zeta)|^2$.
It is important to note that the equation of motion of a Nambu-Goto string, $\ddot{{\bf x}}-{\bf x}''=0$, should not be imposed here.
Deviations may arise from Nambu-Goldstone boson radiation or, in the case of vorton configuration, from a balance between the string tension
and fermion zero-mode energy. Considering a stationary string, $\dot{\bf x}=0$, we have $|{\bf x}'| = 1$ from \cref{eq:conformalG} and
the Lagrangian is further simplified by setting $\sqrt{-\gamma}=|{\bf x}'(\zeta,t)|^2=1$.

The S-matrix for the zero-mode decay process, $\psi^{(0)}\rightarrow q+h$, in the worldsheet formalism is given by,
\begin{equation}
\label{eq:S-matrix-ws}
    i \hat{T}=i\int d^4x\langle p_{f1}p_{f2}|\mathcal{L}_{\rm int}|p\rangle=-i\kappa\, \bar{u}(p_{f1})\eta_R \,
    \int dt \, e^{i(E_f-E)t}\int_0^Ld\zeta e^{-i{\bf p}_f\cdot {\bf x}(\zeta)}e^{i{p}{ \zeta}},
\end{equation}
where $E_f=E_{f1}+E_{f2}$ and ${\bf p}_f={\bf p}_{f1}+{\bf p}_{f2}$ are the total energy and momentum of the final-state quark and Higgs, respectively, 
and $L$ denotes the total length of the string. 
The decay rate is then obtained by integrating over the final 
state phase space,  
dividing by the total time $\Delta {\rm t} $ and string length $L$,
\begin{equation}
    \Gamma(E)= \frac{1}{{\Delta  t}\, L}  \int \frac{d^{3} {\bf p}_{f1}}{(2\pi)^{3}(2E_{f1})}\frac{ d^{3}{\bf p}_{f2}}{(2\pi)^{3}(2E_{f2})} 
          |\hat{T}|^2 \, .
\end{equation}

\paragraph{Circular string loop (vorton) decay \label{sec:2_massless_decay}}

We consider the decay of a circular string loop (vorton) with radius $R$ in a cylindrical coordinate system $(r,\theta,z)$.
The spatial worldsheet coordinate is parameterized as $\zeta=R\theta$, where $\theta\in[0,2\pi)$ (see \cref{fig:coord-loop}). 
With this setup, the $\zeta$-integral in \cref{eq:S-matrix-ws} evaluates to 
\begin{equation}
\label{eq:Izeta}
    I_\zeta\equiv\int_0^{L}d\zeta e^{-i{\bf p}_f\cdot {\bf x}(\zeta)}e^{i{ p}{ \zeta}} 
      = R\int_0^{2\pi}d\theta e^{-iR{\bf p}_{fP}\cdot \hat{\boldsymbol{r}}}e^{iER\theta}=2\pi R \, (-i)^nJ_n(|{\bf p}_{fP}|R),
\end{equation}
where $n\equiv ER\in\mathbb{Z}$ arises from the quantized momentum due to the periodicity of the string loop.
We denote the planar momentum of the final state ${\bf p}_{fP}$ as the momentum component in the string loop plane, with 
magnitude $|{\bf p}_{fP}|=\sqrt{p_{f,x}^2+p_{f,y}^2}$. Due to the azimuthal symmetry, we set the direction of ${\bf p}_{fP}$ be $\hat{x}$.
The S-matrix for this process becomes
\begin{equation}
    i \hat{T}=-i\kappa 2  \pi R\, (2\pi) \delta(E_f-E)(\bar{u}(p_{f1})\eta_R)(-i)^nJ_n(|{\bf p}_{fP}|R) \, , 
\end{equation}
which is consistent with the full $3+1$\,D S-matrix result in \cref{eq:T_scalar}. 
Using the asymptotic expansion of the Bessel function $J_n(|{\bf p}_{fP}|R)$, the decay rate is given by
\begin{equation}
\label{eq:decay_vorton_PS}
\Gamma(E)
 = \int \frac{d^{3}{\bf p}_{f1}}{(2\pi)^{3}(2E_{f1})}\frac{ d^{3} {\bf p}_{f2}}{(2\pi)^{3}(2E_{f2})}
      \kappa^2 (2\pi)\delta(E_f-E)E_{f1}\frac{e^{-\frac{2}{3} n \left(1-\left(\frac{|{\bf p}_{fP}|}{E}\right)^2\right)^{3/2}}}{\sqrt{E^2-|{\bf p}_{fP}|^2}}.
\end{equation}
Neglecting the masses of $q$ and $h$, the decay rate simplifies to 
\begin{equation}
\label{eq:adiabatic-decay-vorton}
   \Gamma(E)\simeq \frac{y_D^2 E}{(2\pi)^4}\left(\frac{E}{m_\psi}\right)^2(ER)^{-2/3} \, ,
\end{equation}
consistent with the results of \cite{Fukuda:2020kym}.  
\\

\noindent
Several comments on the decay rate of a vorton are in order: 
\begin{enumerate}
\item {\it Final-state momentum direction}:
The decay products are emitted along the tangent direction of the string loop.
This is not directly evident from the S-matrix calculation, as the momentum eigenstate of the zero-mode overlaps with the 
entire string. However, the conservation of energy and angular momentum ensures that the total final state momentum total 
almost aligns with that of the zero mode within the uncertainty.

\item {\it Planar momentum range}:
The exponential term in \cref{eq:decay_vorton_PS} confines $p_{fP}$ to values near $E$, with a range
determined by $\sqrt{ 1 - ( |{\bf p}_{fP}| /E )^2 } \lesssim  (ER)^{-1/3}$.\footnote{This is valid when $n(m_1+m_2)^3/E^3\ll1$, in the opposite limit, $\sqrt{ 1 - ( |{\bf p}_{fP}| /E )^2 } \lesssim  (ER)^{-1/2}$.} Combining 
the term $1/{\sqrt{E^2-|{\bf p}_{fP}|^2}}$ in \cref{eq:decay_vorton_PS} from the Bessel 
function expansion with the phase space suppression gives the term $(ER)^{-2/3}$ in the decay rate. The remaining factors 
can be obtained by dimensional analysis.

\item {\it Final-state mass}:
The condition for neglecting the mass of the final state particles is that $ E R (\frac{m_1 + m_2} { E} )^3 \ll 1$. 
When masses are included, the minimum of the exponential term in \cref{eq:decay_vorton_PS} occurs for final states 
with the same velocity ($ \frac{ p_1}{E_1} = \frac{p_2 } {E_2}$), and momenta parallel to the initial momentum. 
This gives the exponential suppression factor, $\exp ( - \frac{2}{3} n (\frac{ m_1 + m_2 }{ E} )^3 )$.   
When $ E R (\frac{m_1 + m_2} { E} )^3 \ll 1$, the mass term can be neglected in the integration.
For electroweak-scale masses, the condition is easily satisfied.

Taking the opposite limit, $ E R (\frac{m_1 + m_2} { E} )^3 \gg 1$, while ensuring $m_1 + m_2 < E$ so that 
energy conservation allows the decay, the decay rate can be estimated as, 
\begin{equation}
\label{eq:decayM12}
   \Gamma(E)\simeq \frac{y_D^2 E}{(2\pi)^4}\left(\frac{E}{m_\psi}\right)^2
   \exp\left(- \frac{2}{3} n  \left(\frac{m_1+m_2}{E}\right)^3\right)\left(\frac{E}{m_1+m_2}\right)^{11/2}(ER)^{-5/2} \, . 
\end{equation}
Details are provided in \cref{app:s-matrix}.

\item {\it Lifetime of a vorton}:
For a vorton generated from a long string at \( T = 1 \, \text{GeV} \) with \( v_a = 10^{10} \, \text{GeV} \) and \( y_\psi = 1 \), the relevant parameters are \( n = \epsilon_F R \sim 10^{15} \) and \( E  = \epsilon_F \sim 2\pi v_a \).  
Here, the energy $ E$ is chosen as the Fermi energy to maximize the decay rate. 
Requiring fermions in the vacuum decay before BBN, with the vacuum decay rate $\Gamma_{\rm vac}(E)=\frac{|y_D|^2}{16\pi}m_\psi$,  
yields a lower bound on the Yukawa coupling, 
\begin{equation}
y_D \ge 5\times 10^{-16}\sqrt{\frac{10^{10}{\rm GeV}}{m_\psi}} \, .
\end{equation}
Using this bound, \cref{eq:adiabatic-decay-vorton} gives the vorton lifetime as,
\begin{equation}
    \tau \sim  7 \times10^{6}\,{\rm s}.
\end{equation}
This corresponds to approximately three months, much shorter than the CMB last-scattering time.

\item{\it Small mixing}:
When $E > m_\psi$, there are no exact bound states and the wavefunction of the approximately bound fermion may have small overlap with free-fermion states, which leads to an additional decay rate.  However, because of the smallness of the mixing, the decay rate computed above should be the dominant one. The overlap of the wavefunction, however, leads to the tunneling of the approximate bound states into free states, which we discuss in \cref{sec:qt}.

\end{enumerate}

We conclude that if the fermion $\psi$ decays into the Standard Model particles outside the string before the BBN, the vorton cannot be stable.
Stable vortons require the extension of the model. For example, if $q$ or $h$ is a 
massive beyond-Standard Model particle, we may evade the BBN bound while making the vorton lifetime exponentially long.
Those new particles $q$ or $h$ should couple to SM particles and decay into them before the BBN. Then the off-shell exchange of $q$ or $h$ mediates the decay of the zero modes into SM particles that is not exponentially suppressed. However, by taking the coupling of $\psi$ to $q$ and $h$ and that of $q$ or $h$ to SM particles both small, the lifetime of $\psi^{(0)}$ can be much longer than the age of the universe.

In \cref{app:GaussianModulation}, we calculate the decay rate of the same process $\psi^{(0)}\rightarrow q+h$ in a static or slowly-moving string segment with a curvature radius $R$. The string segment's configuration is modeled by a Gaussian function, which we denote as ``Gaussian modulation''. The same string configuration is also used when we discuss the non-adiabatic decay process in the following section.

\subsubsection{Non-adiabatic decay: $\psi^{(0)}+\delta\Phi \rightarrow q+h$}

The zero modes $\psi^{(0)}$ can decay into $q$ and $h$ through direct coupling, as described in \cref{sec:psi2qH}.
In addition, variations in the string background introduce another process, $\psi^{(0)}+\delta\Phi\rightarrow \tilde{\psi}\rightarrow q+h$, 
where $\tilde{\psi}$ represents the excited (or continuum) fermion state.   
In this scenario, the zero mode is no longer orthogonal to the excited states in the string background, leading to 
new interaction terms, $ y_\psi \delta \Phi \bar{ \tilde{\psi}}_R \psi_L^{(0)}$, and the associated process
$\psi^{(0)}+\delta\Phi\rightarrow \tilde{\psi}\rightarrow q+h$. 
This phenomenon requires that the background variation occurs rapidly enough for the zero mode to remain unadjusted to the 
new string configuration. This approximation is known as the {\it non-adiabatic approximation}.
Note that even within the non-adiabatic regime, the zero-mode decay persists due to its direct coupling to $q$ and $h$. 
First, we study the parameter space where the non-adiabatic approximation is valid, characterized by a narrow region of string modulation 
radius $R \ll \delta/(\epsilon^3 y_\psi^2) $.

Next, we derive the effective action governing the zero mode coupling to Higgs and quarks under fast background variation $\delta\Phi$.
Finally, we calculate the decay rate using a Gaussian modulation as an example \cite{Ibe:2021ctf}.

\paragraph{Non-adiabatic approximation}
We determine the characteristic timescale and the valid region of the non-adiabatic approximation for zero mode decay, 
following the formalism in \cite{Born:1928cqs,Messiah:1979eg}.

\begin{figure}[htbp]
\centering  
    \begin{subfigure}[b]{0.4\textwidth}
        \centering
        \includegraphics[width=1.1\textwidth]{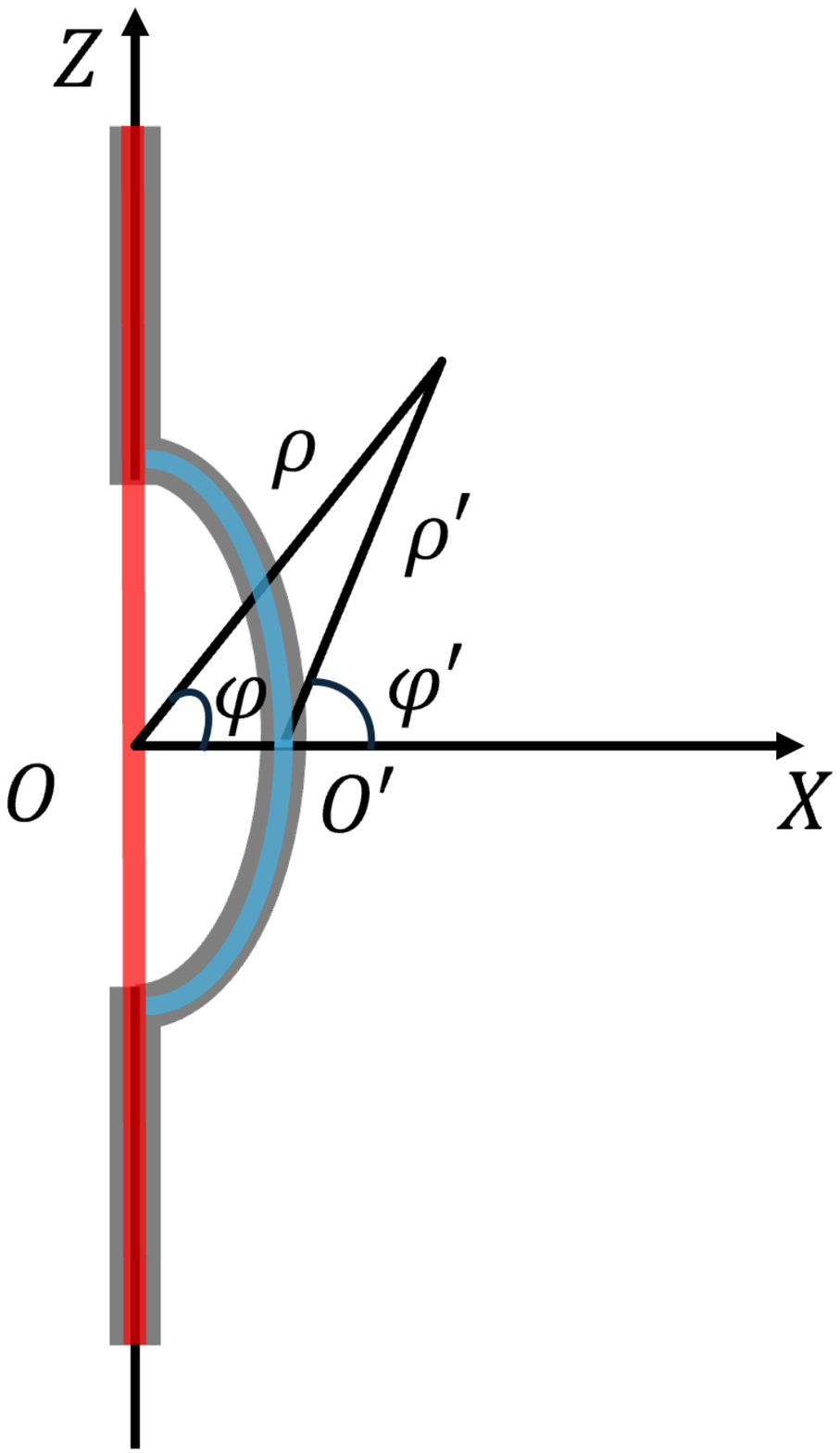}
    \end{subfigure}
    \caption{\label{fig:coord-straight-string} The coordinate setup of an asymptotic straight string with a Gaussian modulation. The string is denoted by a gray line, while the blue and red lines represent the zero mode in the adiabatic and non-adiabatic cases respectively.}
\end{figure}

Consider the zero mode state $| \psi^{(0)} \rangle$ as the initial state of the system. The sudden (non-adiabatic) approximation holds
when the time evolution of the state over time interval $\Delta t$ does not significantly modify the original state, 
\begin{equation}
   {\cal U} ( \Delta t  ) | \psi^{(0)} \rangle   \simeq | \psi^{(0)} \rangle \, ,
\end{equation}
where ${\cal U}(\Delta t)$ is the unitary time evolution operator, approximated as,
\begin{equation}
    \hat{\mathcal{U}}(1)=1-\frac{i}{\hbar}\Delta t\bar{H} \, .
\end{equation}
Here, $\bar{H}$ is the average Hamiltonian over this time interval $\Delta t$.
The approximation requires that the probability of finding the state not in $| \psi^{(0)} \rangle$ is negligible. 
The probability is expressed as 
\begin{equation}
\label{eq:P1}
    {\rm Prob}=
      \langle \psi^{(0)} |
      {\cal U}^\dagger (  \Delta t ) 
         \left( {\bf \hat{I}}-| \psi^{(0)}  \rangle\langle \psi^{(0)} | \right)
        {\cal U}(  \Delta t )  | \psi^{(0)} \rangle \simeq \frac{\Delta t^2}{\hbar^2}(\Delta\bar{H})^2 \ll 1 
\end{equation}
where ${\bf \hat{I}}$ is the identity operator in the Hilbert space, and 
$\Delta\bar{H}$ is the uncertainty of the Hamiltonian $\bar{H}$. 
To satisfy the sudden approximation, ${\rm Prob} \ll 1$, the duration of the process must fulfill
\begin{equation}
    \Delta t\ll \hbar/\Delta\bar{H}.
\end{equation}  
The Hamiltonian uncertainty, $\Delta\bar{H}$, can be estimated from the variation in the background field, $\delta \Phi$,    
\begin{equation}
   \Delta\bar{H} \sim \int d^2 x |\psi^{(0)}|^2 \delta \Phi  \sim m_\psi m_\phi \Delta \rho \, ,
\end{equation}
where the $\Delta \rho = |\boldsymbol{\rho}' - \boldsymbol{\rho}|$ is the displacement of a string segment, illustrated in \cref{fig:coord-straight-string}.
Assuming the displacement is the order of string core size multiplied by a small number $\epsilon$, 
$\Delta \rho \sim \epsilon / m_\phi$, we obtain 
$\Delta\bar{H} \sim \epsilon m_\psi$. 
Thus, the sudden approximation is valid in the region 
$\Delta t \ll \epsilon^{-1}  m_\psi^{-1}$.

The characteristics time scale for string oscillation is $t_{\rm osc}\sim \sqrt{\epsilon \delta R}$, 
where $\delta$ is the string core size, and $R$ represents the curvature radius of the string segment that oscillates. 
This estimation is based on the assumption that the transverse acceleration of the string segment due to its tension is 
approximately $1/R$, and the segment moves a distance of $\epsilon \delta$. 
For the sudden approximation to remain valid, the process time scale must satisfy $\Delta t \ll 1 / (\epsilon m_\psi)$. 
Substituting this condition into the oscillation time scale gives the constraint, 
resulting in 
\begin{equation}
   R \ll \frac{1}{ \epsilon^3 y_\psi^2} \delta \, .
\end{equation}
This result implies a relatively small radius for the non-adiabatic regime. In contrast, the characteristic size of vorton is typically 
much larger, unless the Yukawa coupling, $y_\psi $ or $\epsilon$ are exceptionally small.  
Despite the smallness of this region, it is valuable to explore the effective action and decay rate within the non-adiabatic region, 
as it provides insights into the dynamics under rapid background variations.

\paragraph{Effective action}
In the non-adiabatic (sudden) approximation, 
the zero mode does not have sufficient time to adjust its wavefunction to align with a new energy eigenstate and instead remains 
in its original energy eigenstate during the string modulation process. 
The modulation of string, $\delta\Phi$, effectively scatters with the zero mode via the Yukawa interaction described in 
\cref{eq:L_psi}. The intermediate state $\tilde{\psi}$ subsequently decays into $q$ and $h$. 
This process $\psi^{(0)}+\delta\Phi\rightarrow \tilde {\psi}\rightarrow q+h$ can be also viewed as
the zero $\psi^{(0)}\rightarrow \tilde {\psi}\rightarrow q+h$, treating $\delta\Phi$ as a background field.

We focus on the regime where the characteristic energy scale is smaller than the fermion mass, $ E < m_\psi$. 
In this limit, the zero mode is orthogonal to the continuum states, allowing the fermion to be decomposed 
as $\psi  = \psi^{(0)} + \tilde{\psi}$. This decomposition becomes approximate in the opposite limit, $ E  >  m_\psi$. 
In the large mass limit, the intermediate state $\tilde{\psi}$ can be integrated out. As a caveat, this simplification 
breaks down near the center of the string, an effect we neglect for simplicity.

The fermion field $\psi$ couples to the scalar field $\Phi$ as well as $h$ and $q$  through the Lagrangian,
\begin{equation}
 {\cal L } =  - y_\psi \Phi \, \bar{\psi}_L\psi_R - y_D\bar{q}_L h \psi_R  +{\rm h.c.} \, . 
\end{equation}
Decomposing the fields $\psi  = \psi^{(0)} + \tilde{\psi}$ and $\Phi   =  \Phi_0 + \delta \Phi$, the interaction Lagrangian becomes
\begin{equation}
   {\cal L }_{\rm int} =  
      - y_\psi  \Phi \,  \bar{\tilde \psi}_L \tilde{\psi}_R  
      - y_\psi \delta \Phi \, ( 
        \bar{\tilde \psi}_L\psi_R^{(0)}  
       + \bar{\psi}_L^{(0)} \tilde \psi_R  
      )
      -   y_D\bar{q}_L h  \psi^{(0)}_R  -   y_D\bar{q}_L h  \tilde{\psi}_R     +{\rm h.c.} \, . 
\end{equation}
We integrate out $ \tilde \psi$ by neglecting its kinetic term and 
solving the field equation $ d {\cal L } / d   \bar{\tilde \psi}_L  = 0 $.
Substituting this solution back into the interaction Lagrangian, the resulting effective action is
\begin{equation}
   {\cal L}_{\rm eff}  =  - y_D \bar{q}_L h   \psi^{(0)}_R  +   y_D  \frac{ \delta \Phi}{ \Phi}  \bar{q}_L h  {\psi}_R^{(0)}     +{\rm h.c.}  \,.
\label{eq:Leff1}
\end{equation}
The effective Lagrangian reveals that the scattering process $\psi^{(0)}+\delta\Phi \rightarrow q+h$ can be interpreted as decay, 
but suppressed by $\delta \Phi / \Phi$. Additionally, the zero-mode decay process $\psi^{(0)}  \rightarrow q+h$,
governed by the first term in the effective Lagrangian, always exists
and typically dominates, unless this leading interaction is forbidden or suppressed by symmetry.
The suppression can occur when the initial configuration of the zero mode has low curvature, and the string suddenly transitions to a configuration with larger curvature.

\paragraph{Gaussian modulation example}
Consider a segment of a straight string that is suddenly bent with a Gaussian modulation, represented as
\begin{equation}
\Delta(z)=\epsilon \delta e^{-\frac{z^2}{2\sigma_z^2}} \, , 
\end{equation} 
as illustrated in \cref{fig:coord-straight-string}.
The equivalent curvature radius for this modulation is approximately $R \simeq \frac{\sigma_z^2  }{ \epsilon \delta} $. 
Under the sudden approximation, 
the zero-mode wavefunction remains unchanged during the modulation period. Decay through the first term in \cref{eq:Leff1}
vanishes in a straight string limit, while the second term gives the following matrix element 
\begin{equation}
\label{eq:T_approx}
     i \hat{T} = i y_D (2\pi)\delta(E-E_f)(\bar{u}_q(p_1)P_R\eta)\int d^3x
      \frac{f(\rho')}{f(\rho)}\mathcal{F}(\rho)e^{-i({\bf p}_{f}-{\bf p})\cdot {\bf x}},
\end{equation}
where $p_1$ is the momentum of quark and $ p_{f}$ represents the total momentum of the final states. 
The string background can be expanded as $ f (\rho') \simeq  f( \rho) - f'( \rho) \Delta ( z) \cos \varphi$, 
and we further separate the spatial integration 
into $z$-direction and string loop plane.  
The $z$-integral is given by 
\begin{equation}
\label{eq:I_z_non_ad}
   I_z \equiv \int_{-\infty}^{+\infty} dz\Delta(z)e^{-i(p_{f,z}-p_z)z}  
      =  \sqrt{2\pi} \epsilon\delta  \sigma_z  \, e^{-\frac{\sigma_z^2}{2}(p_{f,z}-p_z)^2} \,.
\end{equation}
In the string loop plane, 
the plane wave $e^{-i({\bf p}_{f}-{\bf p})\cdot {\boldsymbol{\rho}}}$ is expanded into partial waves in polar coordinates. 
The 2D integration becomes 
\begin{eqnarray}
       I_{\rm 2} &\equiv& \int d^2x \mathcal{F}(\rho) e^{-i({\bf p}_{f}-{\bf p})\cdot {\boldsymbol{\rho}}}
    \frac{f'(\rho) }{ f( \rho ) }\cos\varphi
   \nonumber\\
      &=& \int d^2x \mathcal{F}(\rho) \frac{f'(\rho) }{ f( \rho ) }\cos\varphi \, 
            \sum_{n=-\infty}^{+\infty} (-i)^nJ_n(|{\bf p}_{fT}|\rho)e^{-i n(\varphi-\varphi_{T})}
\end{eqnarray}
where $\varphi_{T}$ denotes the azimuthal angle of the total final-state transverse momentum ${\bf p}_{fT}$, i.e.,
$\varphi_{T}=\tan^{-1}\left(p_{f,y}/p_{f,x}\right)$, and ${\bf p}\cdot\boldsymbol{\rho}=0$ since zero-mode momentum is perpendicular to the transverse direction. 
After performing the $\varphi$-integral, the Bessel function order is fixed to $n=\pm1$ due to the presence of 
$\cos\varphi=(e^{i\varphi}+e^{-i\varphi})/2$. 
Expanding the Bessel function $J_1(|{\bf p}_{fT}|\rho)\simeq|{\bf p}_{fT}|\rho$ around zero by assuming $|{\bf p}_{fT}|\rho\sim \rho/R\rightarrow 0^+$ 
in the large curvature radius limit, the radial integral is evaluated using the saddle-point approximation, yielding 
\begin{equation}
    I_{\rm 2}
      \simeq -i p_{f,x}m^{-1}_{\psi} \,  .
\end{equation}
The decay rate is then calculated as,
\begin{equation}
    \Gamma(E)=\int\frac{d^3{\bf p}_1}{(2\pi)^3(2E_1)}\int\frac{d^3{\bf p}_2}{(2\pi)^3(2E_2)}\frac{|\hat{T}|^2}{ \Delta t \,L_{\rm str}}=y_D^2(2\pi\epsilon^2\delta^2\sigma_z)m_\psi^{-2} I_{\rm PS},
\end{equation}
where $L_{\rm str}=\sigma_z$ and the phase space integral is changed by a Gaussian kernel,\footnote{The saddle point approximation requires the final-state momenta to be $-z$ direction, which aligns with the zero-mode momentum (left-moving). So $E_1+p_{1,z}\rightarrow E_1-|p_{1,z}|$ at the saddle point. }
\begin{equation}
\label{eq:I_PS_non_ad}
    I_{\rm PS}\equiv 
      \int\frac{d^3{\bf p}_1}{(2\pi)^3(2E_1)}\int\frac{d^3{\bf p}_2}{(2\pi)^3(2E_2)}(2\pi)\delta(E-E_f)(E_1+p_{1,z})(p_{1,x}+p_{2,x})^2e^{-\frac{\sigma_z^2}{2}(p_{f,z}-p_z)^2}.
\end{equation}
The detailed derivation in \cref{app:GaussianModulation} for massless final states gives
\begin{equation}
    I_{\rm PS}\simeq \frac{1}{2(2\pi)^5}\frac{E^6}{y^4},
\end{equation}
where $y\equiv\sigma_z E$. 
Substituting $\sigma_z$ in terms of its equivalent curvature radius, $R \simeq \frac{\sigma_z^2  }{ \epsilon \delta} $, 
the decay rate becomes  
\begin{equation}
    \Gamma(E)=\frac{1}{2(2\pi)^4}\sqrt{\epsilon} \, y_D^2 E \, 
      \left(\frac{E}{m_\psi}\right)^{2}\left(E R\right)^{- 3/2} ( E \delta)^{1/2} \, .
\label{eq:gamma_nonA}
\end{equation} 
The decay rate is suppressed as $(ER)^{-3/2}$. In the large mass limit,  $ E \delta \lesssim m_\psi / m_\phi$. 

When the adiabatic condition is satisfied, fermions nearly stay the energy eigenstate. There is a small deviation from the energy eigenstate, but the resultant decay rate is much smaller than that in \cref{eq:gamma_nonA}.

Below we comment on the massive final-state case and show that this non-adiabatic decay process is also exponentially suppressed as the consequence of the approximate momentum conservation. 
Let us take the mass of $h$ to be $m_2$.
The phase-space integral is still given by \cref{eq:I_PS_non_ad}. The saddle point of the Gaussian function for the massless case is such that the total final-state momenta aligns with the longitudinal direction and $p_{f,z}=p_z$. However, the saddle point is shifted for the massive case:
\begin{equation}
    (p_{f,z}-p_z)^2_{\rm min}\simeq\left(\sqrt{E^2-m_2^2}-E\right)^2\simeq\frac{m^4_2}{4E^2},
\end{equation}
where $E$ is the zero-mode energy, and we assume $E\gg m_2$ such that the second equality is obtained by Taylor expansion in large $E$ limit. By substituting the minimum value into the Gaussian function, we directly find that the decay rate is exponentially suppressed by
\begin{equation}
\label{eq:nonadiabatic_exp_suppression}
\exp\left(-\frac{\sigma_z^2m_2^4}{8E^2}\right),~~~~\text{ if }~ \frac{\sigma_z^2m_2^4}{8E^2}\gg1,~\text{or }~ m_2\gg \sqrt{\frac{E}{\sigma_z}}.
\end{equation}
For example, by replacing $q$ or $h$ with a new particle with a mass $\sim v_a$, for $E\sim \epsilon_F \sim v_a$, the decay rate is exponentially suppressed as long as $\sigma_z > 1/v_a$. Modulation of the string loop that occurs much after the ${\rm U}(1)$ global symmetry breaking will satisfy $\sigma_z \gg 1/v_a$. 
For example,
a modulation on the string segment can be induced by the Standard Model thermal bath, and the typical width $\sigma_z$ is $1/T$. 
For $T\ll v_a$, the decay rate is exponentially suppressed.
For the thermal modulation, the adiabatic condition is also satisfied, which further suppresses the decay rate.

\subsection{Quantum tunneling: $\psi^{(0)}\rightarrow\psi$ \label{sec:qt}}

As discussed in \cref{sec:circular}, when $E \gtrsim m_\psi$, a stable bound state does not exist in a circular string loop and the tunneling process $\psi^{(0)}\rightarrow\psi$ is induced.
Understanding this process is crucial for determining the maximum Fermi surface of zero modes in the vorton.

We consider a circular string loop with a radius $R$. 
To calculate the tunneling rate, we must integrate over all possible locations where the zero mode $\psi^{(0)}$ transitions into continuum states propagating out of the string. The continuum state exhibits a mass variation along the string radius,  with a correspondingly modified wavefunction. By accounting for the string profile, the complete $\psi^{(0)}\rightarrow\psi$ rate can be evaluated using the S-matrix method described in the previous subsection.
Note, however, that the momentum of the final state is collimated to the direction tangent to the string because of the angular momentum conservation, and the transverse momentum ($r$ and $z$ direction in Fig.~\ref{fig:coord-loop}) is much smaller than the inverse of the size of the string core. The spreading of the wavefunction of the final state in the orthogonal direction is much larger than the core size, and the mass of the final state can be simply taken to be the value outside the string $m_\psi=y_\psi v_a / \sqrt{2}$.

The method to compute the tunneling rate in this scenario is similar to the worldsheet method but with a massive final state. The quantum tunneling rate is proportional to 
\begin{equation}
 \Gamma(E) \propto   
 e^{- \frac{ 2}{3} n (m_\psi/E)^3}.
\end{equation}
Here the mass of the final state introduces exponential suppression, arising from the asymptotic expansion of the Bessel function at large order, 
 $\exp(-\frac{2}{3} n (1-{\bf p}_{fP}^2/E^2)^{3/2})$, where ${\bf p}_{fP}$ is the planar momentum of the final state and $n\sim E R$. 
The minimum value of the exponential occurs at $|{\bf p}_{fP}| = \sqrt{E^2 - m_\psi^2}$.  
Consequently, the tunneling rate depends exponentially on $e^{- \frac{2}{3} n (m_\psi/E)^3}$.
Requiring the exponent to be $\mathcal{O}(1)$ for an unsuppressed rate implies 
$E \sim \sqrt{m_\psi R} m_\psi$. Given $m_\psi R \gg 1$, the zero modes’ energy for unsuppressed tunneling is significantly larger than 
$m_\psi$. As illustrated in \cref{fig:Etot}, this energy corresponds to an “over-shoot” radius much smaller than vorton’s critical radius, 
where the Fermi energy is approximately $m_\psi$.

Non-adiabatic modulation of the string loop can also induce the tunneling process.
The tunneling rate remains exponentially suppressed due to \cref{eq:nonadiabatic_exp_suppression} (with $m_2$ replaced by $m_\psi$) as long as the typical string modulation length $\sigma_z$ is greater than the string core size.

\subsection{Scattering surrounding charged particles\label{sec:accelarated_particle_scattering}}

We study the scattering between charged particles surrounding the string and fermion zero modes.
This process is efficient when the energy of the charged particle satisfies $2|{\bf p}_{e}| \epsilon_F\gtrsim m_\psi^2$. Requiring this process to be exponentially suppressed for the vorton case puts a lower bound on the fermion mass $m_\psi$.

Since the superconducting string carries electric current due to the gauge anomaly inflow, the string induces an electromagnetic field surrounding the string. The gauge potential $A^\mu$ due to the presence of a straight superconducting string along the $z$-axis is
\begin{equation}
A^\mu(\rho) = \left(-\frac{\lambda_Q}{2\pi}\ln\left(\frac{\rho}{R}\right),0,0,-\frac{\lambda_Q}{2\pi}\ln\left(\frac{\rho}{R}\right)\right),
\end{equation}
where $R$ is the IR cutoff in the transverse direction of the string (for vorton, $R$ is the vorton radius) and $\lambda_Q$ is the linear charge density of the string.
The corresponding electric and magnetic fields are
\begin{equation}
\label{eq:string-E-B}
    {\bf E}_\rho = -\frac{\partial A^0}{\partial \rho}\hat{\rho}=\frac{\lambda_Q}{2\pi}\frac{\hat{\rho}}{\rho},~{\bf B}_\varphi = -\frac{\partial A^3}{\partial \rho}\hat{\varphi}=\frac{\lambda_Q}{2\pi}\frac{\hat{\varphi}}{\rho}.
\end{equation}
The electric field and magnetic field in the vicinity of the superconducting string can be sufficiently large such that the pair production of charged particles and antiparticles is possible~\cite{Schwinger:1951nm}.
The energy spectrum of pair-produced electrons scales as $\exp(-\pi {\bf p}_e^2/|e{\bf E}_\rho|)$ in the near-homogeneous electric field close to the string core.
For sufficiently small $\rho$, 
pair-produced electrons 
can have an energy above $m_\psi^2/\epsilon_F$
to efficiently scatter off fermion zero modes.
Below we derive the condition for the efficient scattering not to occur.

The charge density of a vorton is given by \cref{eq:vorton_Lc},
\begin{equation}
\label{eq:lambda_Qc}
    \lambda_{Q,\rm c} \equiv \frac{Q}{L_c}= ev_a {L}_n,
\end{equation}
where ${L}_n\simeq\sqrt{1+\ln(L_c/(2\pi\delta))}$ contains the logarithmic term from the string tension. Adding back the logarithmic term to the charge density is important for the following discussion on the constraint of the Yukawa coupling $y_\psi$, since we will show that the fermion mass should satisfy $m_\psi^2\gg 2|{\bf p}_e|\epsilon_F$ to make the vorton stable, where both electron energy $|{\bf p}_e|$ and zero-mode Fermi energy are enhanced by $L_n$.

The UV boundary $\rho_{\rm UV}$ above which \cref{eq:string-E-B}
is valid depends on the zero-mode transverse wavefunction spreading, which is given in \cref{eq:zm_spreading}: $ \rho_{\rm UV}\simeq m_\psi^{-1}$ when $m_\phi\gg m_\psi$, and $  \rho_{\rm UV}\simeq (m_\phi m_\psi)^{-1/2}$ when $m_\psi\gg m_\phi$. Let us discuss these two $\rho_{\rm UV}$ regimes:

1. $\rho_{\rm UV}\simeq m_\psi^{-1}$: The largest electric field is $|{\bf E}_{\rho, \rm max}|=\lambda_Q m_\psi/(2\pi)$. The typical pair-produced electron momentum is thus $|{\bf p}_e|\simeq \sqrt{|e{\bf E}_{\rho, \rm max}|}=\sqrt{e\lambda_Qm_\psi/(2\pi)}$. By plugging into the vorton charge density \cref{eq:lambda_Qc}, we find that $|{\bf p}_e|\simeq\sqrt{e^2v_aL_nm_\psi/(2\pi)}$. 

2. $\rho_{\rm UV}\simeq(m_\phi m_\psi)^{-1/2}$: The largest electric field is $|{\bf E}_{\rho, \rm max}|=\lambda_Q (m_\phi m_\psi)^{1/2}/(2\pi)$. The typical pair-produced electron momentum is thus $|{\bf p}_e|\simeq \sqrt{|e{\bf E}_{\rho, \rm max}|}=\sqrt{e\lambda_Q(m_\phi \textit{}m_\psi)^{1/2}/(2\pi)}$. Similarly, we find that $|{\bf p}_e|\simeq\sqrt{e^2v_aL_n(m_\phi m_\psi)^{1/2}/(2\pi)}$. 

The relevant scattering process is $e^-+\psi^{(0)}\rightarrow e^-+\psi$.
The scattering rate of this process can be evaluated using the worldsheet formalism, and it is straightforward to obtain the exponential dependence:
\begin{equation}
\label{eq:exp_scat}
 |\hat T|^2 \propto \exp\left(-\frac{2}{3}n\left(1-\left(\frac{|{\bf p}_{fP}-{\bf p}_{eP}|}{E}\right)^2\right)^{3/2}\right),
\end{equation}
where $n=E R$, with $E$ the zero-mode energy and $R$ the vorton radius, ${\bf p}_{eP}$ is the incoming electron planar momentum, and ${\bf p}_{fP}$ is the total final-state planar momentum.
Evaluating the saddle point of the exponential function, we find that the final-state momentum peaks at the massive particle $\psi$ such that 
\begin{equation}
   \left(\frac{|{\bf p}_{fP}-{\bf p}_{eP}|}{E}\right)^2_{\rm max}\simeq \left(\frac{|{\bf p}_{fP}|}{E}\right)^2_{\rm max}-\left(\frac{2{\bf p}_{fP}\cdot {\bf p}_{eP}}{E^2}\right)_{\rm min}\simeq1-\left(\frac{m_\psi}{E}\right)^2+2\frac{|{\bf p}_{e}|}{E},
\end{equation}
where we neglect the subleading terms $\mathcal{O}(|{\bf p}_{e}|^2/E^2)$ and $\mathcal{O}(m_\psi |{\bf p}_{e}|/E^2)$ and approximate $E+E_e\simeq  E$, with $E_e\simeq |{\bf p}_{e}|$. Considering the Fermi energy of the zero mode $ \epsilon_F\sim 2\pi v_aL_n$, we find that the saddle point is shifted by a small value $2 |{\bf p}_{e}| /\epsilon_F$ compared to the massive 2-body decay process discussed in \cref{app:s-matrix}. By plugging the maximum value back to \cref{eq:exp_scat}, the exponential suppression shows up if
\begin{equation}
   \frac{2}{3}n\left(\left(\frac{m_\psi}{\epsilon_F}\right)^2-2\frac{|{\bf p}_{e}|}{\epsilon_F}\right)^{3/2}\gg 1,
\end{equation}
where we take $E=\epsilon_F$.
The exponential suppression requires $m_\psi^2\gtrsim 2 |{\bf p}_{e}| \epsilon_F$.
Note that the distribution of the electron energy has an exponential tail above $\sqrt{e|{\bf E}_{\rho, \rm max}}|$. Therefore, to exponentially suppress the scattering rate, $\sqrt{e|{\bf E}_{\rho, \rm max}}|$ should be much smaller than $m_{\psi}^2/\epsilon_F$. 
Applying this constraint to the two $\rho_{\rm UV}$ regimes, we find: 1. For $\rho_{\rm UV}\simeq m_\psi^{-1}$,  $m_\psi\gg (8\pi e^2)^{1/3}L_nv_a$. 2. For $\rho_{\rm UV}\simeq (m_\phi m_\psi)^{-1/2}$,  $ m_\psi^{7/2}\gg 8\pi e^{2}m_\phi^{1/2}(L_n v_a)^3$.

The same result applies to the QCD case by replacing the gauge coupling $e$ with the strong coupling $g_s$. 
For $m_\phi \gg m_\psi$, the exponential suppression requires $y_\psi \gg 4\pi$, which violates the unitarity. For $m_\phi \ll m_\psi$, the exponential suppression can be achieved by small $m_\phi$. Small $m_\phi$ is natural in supersymmetric theory where $m_\phi$ can arise from supersymmetry breaking.

The high-energetic electrons can also induce the process $e + \psi^{(0)} \rightarrow e+ q+ h$. 
In the massless limit for all the final-state particles, the scattering rate is similar to the decay rate of the zero-mode three-body decay process, since the incoming electron momentum is subdominant compared to the zero-mode momentum, and the saddle point is close to the one obtained in the massless three-body decay, which gives a power-law suppression and a massive propagator suppression $\Gamma_{\rm scat}\propto (ER)^{-4/3}\times m_\psi^{-4}$. 
The rate is not exponentially suppressed and can destabilize the vorton. To avoid that, one need $2|{\bf p}_e| \epsilon_F \ll m_{1,2}^2$, which requires $m_{1,2}^4\gg(8\pi e^2)(L_nv_a)^3m_\psi$ for $m_\phi\gg m_\psi$, and $m_{1,2}^{4}\gg (8\pi e^2)(L_nv_a)^3 (m_\phi m_\psi)^{1/2}$ for $m_\psi\gg m_\phi$. Here $m_{1,2}$ are masses of $q$ and $h$. The inequality is violated for the SM particles, so the KSVZ fermion should couple to massive beyond-SM particles for it to decay much before the BBN outside the string.

As we explain in Sec.~\ref{sec:friction}, the electrons produced by the electric field screen the charge on the string and weaken the electric field outside the string. Still, the maximal electric field is the same as those used above and the condition for the exponential suppression is unchanged. The screening makes $\rho$ at which the electric field takes the maximal value smaller, but we find that the de Broglie wavelength is still of the same order as $\rho$ and the pair production occurs.

The pair-produced electrons and/or electrons in the thermal bath can be accelerated by the electric field. However, the acceleration time scale is at the shortest $\rho_{\rm UV}$, which is much longer than the time scale for the pair production to occur, $|{\bf p}_e|^{-1}$, and the screening occurs first before the electrons are accelerated. As a result, the effect of acceleration is negligible.

\subsection{Other processes\label{sec:other_process}}
Compared with the non-superconducting cosmic string, the existence of fermion zero modes along the superconducting string introduces a novel channel for scattering processes between the string and the Standard Model particles, which can play an important role in determining the stability of vortons and the string dynamics of the string network in the early universe. First, 
we show that when the thermal plasma temperature is above the freeze-out temperature $T>T_{\rm scat}$, the inelastic scattering between fermion zero modes and Standard Model particles can be a dominant process to dissipate fermion zero modes and thus decrease the charge of the superconducting string and make the vorton unstable. When the temperature drops below $T_{\rm scat}$, this process becomes irrelevant. After that, we analyze the friction provided by the scattering between the string-induced (color-)magnetic field and the Standard Model charged particles in the surrounding plasma, which delays the string network entering the scaling regime after $T_{\rm scaling}$.

\subsubsection{Thermal plasma inelastic scattering \label{sec:plamsa_inelastic_scattering}}
The fermion zero mode can inelastically scatter with the Standard Model particles in the surrounding plasma. This scattering is relevant when the scattering rate is greater than the Hubble parameter: $\Gamma_{\rm scat}(T)>H(T)$.
Thermal gluons can induce the process $g+ \psi^{(0)} \rightarrow g + \psi$, but the rate is exponentially suppressed for $T \ll m_\psi$. We instead consider the coupling of $\psi$ with
 a SM quark $q$ and a Higgs $h$. (If one of them are a massive BSM particle, the scattering rate discussed below is exponentially suppressed at temperatures below its mass.)
By considering the minimal model, the relevant scattering processes include $\psi^{(0)}+g\rightarrow q+h$, $\psi^{(0)}+\bar{q}\rightarrow g+h$, and $\psi^{(0)}+h\rightarrow q+g$.

We explicitly derive the scattering rate of the process $\psi^{(0)}(p)+h(p_a)\rightarrow q(p_1)+g(p_2)$ ($s$-channel with a SM quark as a mediator) as an example, where $p,\, p_a,\, p_1, \,p_2$ are the corresponding four momenta. For a straight-string configuration, the scattering cross-section for the $s$-channel is
\begin{equation}
    \sigma=\frac{1}{(2E_a)|\Delta v|}\int\frac{d^3{\bf p}_1}{(2\pi)^3(2E_1)}\int\frac{d^3{\bf p}_2}{(2\pi)^3(2E_2)}\frac{|\hat{T}|^2}{\Delta t \,L},
\end{equation}
where $|\Delta v|$ is the relative velocity between the incoming Higgs and zero mode, and the normalized S-matrix square is
\begin{eqnarray}
    \frac{|\hat{T}|^2}{\Delta t \,L}=(2\pi)^4\delta(E_f-E_i)\delta(p_{f,z}-p_{i,z})y_D^2g_s^2\frac{4E}{m_\psi^{2}s^2}\left[2(E_i+p_{i,z})(p_1\cdot p_i)-(E_1+p_{1,z})p_i^2\right],\,~~~~~
\end{eqnarray}
where we defined the center-of-mass energy $s\equiv(p+p_a)^2=m_h^2+2p\cdot p_a\simeq 2EE_a(1+\cos\theta_a)$, $\theta_a$ is the scattering angle between the Higgs momentum and zero-mode momentum, and we drop the Higgs mass term because $E_a\gg m_h$ for the thermal Higgs at high temperatures, $p_i^\mu\equiv(E_i,{\bf p}_{i})=p^\mu+p_a^\mu$, $p_f^\mu\equiv(E_f,{\bf p}_{f})=p_1^\mu+p_2^\mu$ is the total initial-state and final-state 4-momentum respectively, and $g_s$ is the strong coupling constant. By doing the phase-space integral, the cross-section is thus equal to
\begin{equation}
    \sigma = \frac{2}{3}y_D^2g_s^2\frac{E_a^3}{|\Delta v|m_\psi^{2}s^2}(3E+2E_a-3E_a\cos\theta_a)(1+\cos\theta_a)\cos^6\left(\frac{\theta_a}{2}\right).
\end{equation}
The scattering rate can be estimated by $ \Gamma_{\rm scat}=n_a\langle \sigma |\Delta v|\rangle$, where $n_a=\frac{\zeta(3)T^3}{\pi^2}$ is the number density of Higgs in the thermal plasma, and the velocity average cross-section is,
\begin{eqnarray}
    \langle \sigma |\Delta v|\rangle=\frac{1}{2}\int_{-1}^{1} d\cos\theta_a \sigma|\Delta v|  \simeq\frac{1}{6}y_D^2g_s^2\frac{E_a(E+E_a/2)}{m_\psi^{2}E^2}
    \simeq \frac{1}{6}m_\psi^{-2}y_D^2g_s^2\frac{T}{E},
\end{eqnarray}
where we consider the maximum zero mode energy $E=\epsilon_F\gg T$, and the typical momentum of the thermal plasma is $|{\bf p}_a|\sim E_a\sim T$. 

The scattering rate is therefore given by 
\begin{equation}
    \Gamma_{\rm scat}=\frac{\zeta(3)}{6\pi^2}y_D^2g_s^2\frac{T^4}{Em_\psi^{2}}.
\end{equation}
By setting $\Gamma_{\rm scat}\le H=\sqrt{\frac{4\pi^3 g_*}{45}}\frac{T^2}{M_{pl}}$, we obtain the freeze-out temperature for the scattering process,
\begin{equation}
    T_{\rm scat}=8.3\times10^5\,{\rm GeV}\,|y_D|^{-1}g_s^{-1}\frac{m_\psi}{10^{10}\,{\rm GeV}}\sqrt{\frac{E}{10^{10}\,{\rm GeV}}}.
\end{equation}
Therefore, in the early universe when the temperature is higher than the freeze-out temperature $T>T_{\rm scat}$, the inelastic scattering between fermion zero modes and Standard Model Higgs (and gluons, quarks) is an efficient and dominant dissipation channel.

\subsubsection{Friction and scaling regime \label{sec:friction}}

Since the electric current induces a magnetic field around the superconducting string, it is also important to consider friction caused by the elastic scattering between the string-induced magnetic field and relativistic Standard Model particles (e.g., electrons, quarks, and so on) in the thermal plasma. The above argument also applies to the color current along the superconducting string.

According to \cref{sec:accelarated_particle_scattering}, pair production of charged particles happens around the superconducting string. Taking the electron-positron pair in the string-induced electric field given by \cref{eq:string-E-B} as an example, if $\lambda_Q$ is positive, the electron produced by the pair production will be accelerated towards the string core, while the positron will be repelled away from the string.
Since the closer to the string, the shorter the pair-production time scale, electrons are first created near the string, screening the charge. 
The pair productions with the screened charge density continue outward. The resultant effective linear charge density $\lambda_Q^{\rm eff}(\rho)$ can be obtained by minimizing the energy $\epsilon_e$ of a bound-state relativistic electron in the 2D Coulomb potential, where  
\begin{equation}
    \epsilon_e \sim (\rho^{-2}+m_e^2)^{1/2}+e\lambda_Q^{\rm eff}(\rho)\ln\left(\frac{\rho}{R}\right).
\end{equation} 
The effective linear charge density is
\begin{equation}
    \lambda_Q^{\rm eff}\simeq \frac{1}{e\rho}, ~~~{\rm when }~\lambda_Q^{\rm eff}\gtrsim m_e/e.
\end{equation}
The screened electric field produced by the effective linear charge density is thus given by $|{\bf E}_{{\rm eff},\rho}|\simeq1/(2\pi e\rho^2)$.
The screening also happens for the $SU(3)_c$ gauge field.

Due to this screening effect, the induced magnetic field will scale as $|{\bf B}_\varphi|=\frac{1}{2\pi e \rho^2}\simeq \rho^{-2}$. Compared with the thermal plasma-induced magnetic field $|{\bf B}_{\rm plasma}|\sim T^2$, we determine the boundary of the magnetic field in the radial distance from the string center $\rho_b\sim T^{-1}$ such that $|{\bf B}_\varphi(\rho_b)|\sim |{\bf B}_{\rm plasma}|$. Since the scattering between the string-induced magnetic field and electrons in the thermal plasma is only relevant within $\rho_b$, the largest scattering cross-section per unit length by dimensionality analysis is thus 
\begin{equation}
    \sigma_{l,{\rm EM}} = \frac{\alpha_{\rm EM}}{T}.
\end{equation}
We can extend this result to the $\rm QCD$ interaction by replacing $\alpha_{\rm EM}$ with $\alpha_{\rm s}$. In this case, the color current of KSVZ fermion zero modes induces a color magnetic field around the long string, and the quark-gluon plasma can be scattered with the color magnetic field. The color sum gives an $\mathcal{O}(1)$ prefactor. 

The friction per unit length due to plasma scattering can be directly obtained 
\begin{equation}
    f_{{\rm f},i} =\sum_a n_{a,i}\sigma_{l,i} v_s \Delta p\sim \mathcal{C}_igT^3 \frac{\alpha_i}{T}v_sT=\mathcal{C}_ig\alpha_i T^3v_s, ~~ i={\rm EM},\,{\rm s},
\end{equation}
where $n_{a,i}$ is the number density of thermal plasma particles of species $a$ that interact with the string-induced color- or electromagnetic field, $g=2$ is the degree of freedom of the thermal plasma particle, $\mathcal{C}_i=\zeta(3)/\pi^2\times N_{b,i}+3\zeta(3)/(4\pi^2)\times N_{f,i}$ for $N_{b,i}$ types of bosonic particles and $N_{f,i}$ types of fermionic particles, $v_s$ is the string velocity, $\Delta p\sim T \Delta v$ is the typical momentum transfer, $\Delta v={\rm max}(v_s, v_{\rm th})$ characterizes relative velocity between the string segment and the plasma, with $v_{\rm th}$ the thermal velocity of plasma. The thermal plasma is ultra-relativistic, so $\Delta v\sim 1$. The friction relaxation time can be estimated by the time scale when the friction dissipates all the string kinetic energy,
\begin{equation}
\label{eq:tau_f}
    \tau_{\rm f}\equiv \frac{\mu v_s^2}{\dot{\epsilon}}=\frac{\mu v_s^2}{ f_{{\rm f},i} v_s}\sim \frac{\mu }{\mathcal{C}_ig\alpha_i T^3},
\end{equation}
where $\mu=\pi v_{a}^2\ln(L/\delta)$ is the string tension, $\dot{\epsilon}=f_{{\rm f},i} v_s$ is the energy loss of string due to friction. Consider a long string with curvature $\frac{1}{R}\sim\frac{1}{L}$, the string velocity is accelerated by force due to curvature during $\tau_{\rm f}$,
\begin{equation}
\label{eq:v_s}
    v_s\sim\frac{\tau_{\rm f}}{L}.
\end{equation}
When the string velocity $v_s\sim1$ and the string length $L\sim H^{-1}$, the long string enters the scaling regime\footnote{These two conditions are related because the distance that a string segment can travel in one Hubble time determines the correlation length of strings in the scaling regime. Since in the scaling regime, the correlation length is about $H^{-1}$, therefore the string velocity should be about $1$.}. By combining \cref{eq:tau_f} and \cref{eq:v_s}, one can obtain the temperature when the string enters the scaling regime,
\begin{equation}
    T_{\rm scaling}\sim \sqrt{\frac{4\pi^3g_*(T)}{45}}\frac{\pi v_a^2\ln(v_a H^{-1}/\sqrt{2})}{\mathcal{C}_{\rm s}g\alpha_{\rm s} M_{\rm pl}}\sim 4.9\times 10^{4}\,{\rm GeV}\left(\frac{v_a}{10^{10}{\rm GeV}}\right)^2,~~{\rm when~ }T_{\rm scaling}\gtrsim1\,{\rm GeV},
\end{equation}
where we set the self-coupling of the PQ scalar field as $\lambda=1$, so the core size of the string $\delta=\sqrt{2}v_{a}^{-1}$, we neglect the friction due to EM interaction between the plasma and string-induced magnetic field since it is subdominant compared to the one due to strong interaction, and we set the factor $\mathcal{C}_{\rm s}g\approx 6.3$ at $T\gtrsim1\,{\rm GeV}$.

\section{Conclusion\label{sec:conclusion}}

In this paper, we study the stability of superconducting global string loops stabilized by fermions trapped inside the loops (i.e., zero modes), named {\it vortons}, at both classical and quantum levels. We find that the vortons can be stable.

We first investigate how vortons shrink.
We perform the lattice simulation for a circular string loop with various initial loop sizes. We measure its asymptotic Lorentz factor before its first collapse, and we find the Lorentz factor is $\mathcal{O}(1)$ for the loop sizes used in the simulation, which shows that the kinetic energy of a circular string loop is comparable to its mass. The result indicates that Nambu-Goldstone boson radiation is efficient in dissipating the string energy, so not all the initial string mass converts to the string kinetic energy. The loop shrinkage will be stopped by the Fermi pressure of the zero-modes and the string loop will not overshoot to eject zero-mode fermions. 

We next investigate the classical dynamics of the zero modes and find that the energy of the fermions $E$ can be as large as $m_\psi \sqrt{m_\psi R}$, where $R$ is the radius of the vorton and $m_\psi$ is the mass of the fermion outside strings. Therefore, for a sufficiently large radius of the vorton, trapped fermions with the Fermi energy do not classically jump out of the vortons.

The possible decay of the zero modes plays an important role in determining the stability of vortons at the quantum level. We introduced the string worldsheet formalism to compute the decay rate of the zero modes into other particles for a general string configuration. Also, we mediate the disagreement in the literature \cite{Fukuda:2020kym,Ibe:2021ctf} on the zero-mode decay rate, by pointing out that
the latter result is valid when the string shape changes non-adiabatically. Furthermore,
we find that if some of the final-state particles of the decay is massive, the decay rate is exponentially suppressed.

When the energy of zero modes $E$ is above $m_\psi$, the zero modes can quantum mechanically tunnel to outside of vortons.
The quantum tunneling rate of zero modes in a vorton is calculated by the S-matrix of
the transition of zero modes into free-states outside the vorton.
We find that the rate is exponentially suppressed when $E \ll m_\psi \sqrt{m_\psi R} $.
This result is consistent with the classical analysis.

Charged particles pair-produced by the electromagnetic field around the superconducting string can scatter efficiently with fermion zero modes along the string when $ 2|{\bf p}_e| \epsilon_F \lesssim 
m_\psi^2$. We find that the scattering rate depends on the spreading of the zero-mode wavefunction
that depends on the hierarchy between $m_\phi$ and $m_\psi$.
The rate can be exponentially suppressed for 
$m_\phi \ll m_\psi^7/v_a^6$.

In this paper, we focused on global strings, but our analysis on the decay and tunneling of fermions trapped inside the string loop is applicable also to local strings. 

Our finding on the stability of vortons set the stage for the study of the phenomenology of stable vortons in both the early and late universe. More intriguingly, stable vortons can serve as cold dark matter candidates, which is the main topic of our follow-up paper \cite{KXXF}.

\section*{Acknowledgments}
We thank Junwu Huang, Sergey Sibiryakov, Pierre Sikivie and Tanmay Vachaspati for their insightful and useful discussion,
Junwu Huang for pointing out the scattering with accelerated particles, Pierre Sikivie for pointing out the possible instability of vortons by the overshoot of the shrinkage of loops, and Tanmay Vachaspati for commenting on the draft.
X.N., W.X., and F.Y.~are supported in part by the U.S.~Department of Energy under grant DE-SC0022148 at the University of Florida. K.H.~is supported by the U.S.~Department of Energy under grant DE-SC0025242 at the University of Chicago, a Grant-in-Aid for Scientific Research from the Ministry of Education, Culture, Sports, Science, and Technology (MEXT), Japan (20H01895), and by World Premier International Research Center Initiative (WPI), MEXT, Japan (Kavli IPMU).

\appendix
\section{Abrikosov ansatz\label{sec:aa}}
In this appendix, we construct the circular loop configuration of radius $R$. Up to the correction of order $\mathcal{O}(\delta/R)$, the phase angle of the complex scalar field can be estimated as the summation of the phases of the vortex-anti vortex pair, and the profile function is modified by $\mathcal{O}(\delta/R)$.

Let us denote the exact solution to the complex scalar field of the circular string loop as $\Phi$. For convenience, we define the dimensionless Abrikosov ansatz $\hat{\Phi}_0$ \cite{Abrikosov:1956sx}, and the dimensionless residual field $\hat\phi$.\footnote{Here, all the fields are dimensionless with the normalization by VEV $\frac{v_a}{\sqrt{2}}$, and all the lengths are dimensionless with the normalization by string core size $(\frac{\sqrt{\lambda}v_a}{\sqrt{2}})^{-1}$. All these dimensionless fields and lengths are labelled with "$\: \hat{ } \:$".} They are related by
\begin{equation}
\label{eq:def}
    \Phi=\frac{v_a}{\sqrt{2}}\hat{\Phi}=\frac{v_a}{\sqrt{2}}(\hat{\Phi}_0+\hat{\phi}),
\end{equation}
where 
\begin{equation}
    \hat{\Phi}_0({\bf \hat{r}})=\hat{\Phi}_1({\bf \hat{r}-\hat{r}_1})\hat{\Phi}_2^*({\bf \hat{r}-\hat{r}_2})
\end{equation}
is the Abrikosov ansatz, ${\bf \hat{r}_1,\hat{r}_2}$ are the locations(normalized by $\delta$) of the vortex-antivortex pair. $\hat{\Phi}_1(\hat{\rho},\varphi)=\hat{\Phi}_2(\hat{\rho},\varphi)=f(\hat{\rho})e^{i\varphi}$ are the usual global string solution. 
The field equations for $\hat{\Phi}$ is
\begin{equation}
\label{eq:eom1}\partial_{\hat{\mu}}\partial^{\hat{\mu}}\hat{\Phi}+\frac{1}{2}\hat{\Phi}(\hat{\Phi}^* \hat{\Phi}-1)=0.
\end{equation}
Substitute \cref{eq:def} into \cref{eq:eom1}, we get 
\begin{eqnarray}
\label{eq:eom2}
\partial_{\hat{\mu}}\partial^{\hat{\mu}}(\hat{\Phi}_0+\hat{\phi})+\frac{1}{2}(\hat{\Phi}_0+\hat{\phi})\left[(\hat{\Phi}_0+\hat{\phi})^* (\hat{\Phi}_0+\hat{\phi})-1\right]=0,\nonumber\\
\Rightarrow\partial_{\hat{\mu}}\partial^{\hat{\mu}}\hat{\Phi}_0+\frac{1}{2}\hat{\Phi}_0(\hat{\Phi}_0^*\hat{\Phi}_0-1)+L_{\hat{\phi}}=0,
\end{eqnarray}
where $L_{\hat{\phi}}=\partial_{\hat{\mu}}\partial^{\hat{\mu}}\hat{\phi}+\frac{1}{2}(\hat{\Phi}_0^2\hat{\phi}^*+2|\hat{\Phi}_0|^2\hat{\phi}-\hat{\phi})+\mathcal{O}(\hat\phi^2)$. Let's first estimate the order of magnitude of the residual field by the equation of motion in eq.\,(\ref{eq:eom2}). By using the asymptotic behavior at a large distance of the global string solution,
\begin{equation}
    f(\hat{\rho})=1-\mathcal{O}(\hat{\rho}^{-2}),~~\hat{\rho}\rightarrow\infty,
\end{equation}
substituting the ansatz into eq.\,(\ref{eq:eom2}),
 we find that near $\hat{\Phi}_1$ string core 
\begin{equation}
    \hat\Phi_1L(\hat\Phi_2^*)+\hat\Phi_2^*L(\hat\Phi_1)+L_{\hat\phi}=\mathcal{O}(\hat{d}^{-1}),
\end{equation}
where $L(\hat{\Phi}_i)$ is the Euler-Lagrange equation of scalar field $\hat{\Phi}_i$ in the straight string configuration, which has the same expression as \cref{eq:eom1}, $\hat{d}=\frac{|{\bf r_1-r_2}|}{\delta}=\frac{2R}{\delta}\gg1$ is the dimensionless separation of the vortex-antivortex pair. Since $L(\hat{\Phi}_i)=0$ for the global string configuration, we find that $L_{\hat\phi}\sim\mathcal{O}(\hat{d}^{-1})$. Furthermore, one can show that the residual field $\hat{\phi}\sim\mathcal{O}(\hat{d}^{-1})$. The $\mathcal{O}(\hat{d}^{-1})$ correction is from the azimuthal angle derivative in the term $2\partial_{\hat{\mu}}\hat\Phi_2^*\partial^{\hat{\mu}}\hat\Phi_1$.

Next, we want to find the explicit solution to \cref{eq:eom1} with correction up to $\mathcal{O}(\delta/R)$ using the iteration method by considering only $\mathcal{O}(\delta/R)$ term when solving the equation of motion, \cref{eq:eom1}.

Up to $\mathcal{O}(\hat{d}^{-1})$ correction, the equation of motion becomes
\begin{equation}
\label{eq:iteration}
    L_{\hat{\phi}}^{(1)}+\frac{2}{\hat{\rho}\hat{ d}}\hat{\Phi}_1e^{-i\varphi_2}=0,
\end{equation}
where $\varphi_2$ is the local winding angle of the $\hat{\Phi}_2$ vortex, and
\begin{equation}
    L_{\hat{\phi}}^{(1)}=\frac{\lambda}{2}\hat{\Phi}_0^2\hat{\phi}^*+\lambda(|\hat{\Phi}_0|^2-\frac{1}{2})\hat\phi
\end{equation}
are the leading order terms. We use the polar coordinate system with the origin  at the center of vortex 1. The ansatz field can be evaluated by using the global string solution and also the asymptotic behavior at a large distance, 
\begin{equation}
    |\hat{\Phi}_0|^2=|\hat\Phi_1|^2|\hat\Phi_2^*|^2=f(\hat{\rho})^2(1-2\mathcal{O}(\hat{d}^{-2})),
\end{equation}
\begin{equation}
    \hat{\Phi}_0^2=\hat{\Phi}_1^2\hat{\Phi}_2^{*2}=f(\hat{\rho})^2e^{i2(\varphi_1-\varphi_2)}(1-2\mathcal{O}(\hat{d}^{-2})).
\end{equation}
By substituting these back into \cref{eq:iteration}, we have
\begin{equation}
\partial_{\hat{\mu}}\partial^{\hat{\mu}}\hat{\phi}+\frac{1}{2}f(\hat{\rho})^2e^{i2(\varphi_1-\varphi_2)}\hat{\phi}^*+(f(\hat{\rho})^2-\frac{1}{2})\hat\phi=-\frac{2}{\hat{\rho} \hat{d}}f(\hat{\rho})e^{i(\varphi_1-\varphi_2)}.
\end{equation}
To match the phase of each term above, we need $\hat{\phi}\propto e^{i(\varphi_1-\varphi_2)}$. Near the $\hat{\Phi}_1$ vortex when $\hat{\rho}\rightarrow0$, $f(\hat{\rho})\propto \hat{\rho}$, we get the solution of leading order of $\mathcal{O}(\hat{d}^{-1})$ \footnote{By adding back the dimension of the distance, we can get back the dimension of the field.}
\begin{equation}
    \hat{\phi}=\frac{\mathcal{O}(1)}{\hat{\rho}\hat{ d}}f(\hat{\rho})e^{i(\varphi_1-\varphi_2)}.
\end{equation}
Away from the $\hat{\Phi}_1$ vortex, $f(\hat{\rho})^2\rightarrow 1$, we get instead
the solution is given by
\begin{equation}
    \hat{\phi}=\frac{\mathcal{O}(1)}{\hat{\rho} \hat{d}}e^{i(\varphi_1-\varphi_2)}.
\end{equation}
Therefore the solution to the total complex scalar field up to $\mathcal{O}(\frac{\delta}{R})$ correction is, 
\begin{equation}
\label{eq:Phiexact}
    \Phi=\frac{v_a}{\sqrt{2}}f(\rho)\left(1+\mathcal{O}(1)\frac{\delta^2}{\rho d}\right)e^{i(\varphi_1-\varphi_2)}.
\end{equation}

\section{Anomaly inflow from the view of axion-QED \label{app:axion-QED}}
Given the axion-photon interaction Lagrangian in \cref{eq:L_axion} and the photon's kinetic term,
\begin{equation}
    \mathcal{L}=-\frac{1}{4}F_{\mu\nu}F^{\mu\nu}+\frac{E_{\mathcal{A}}}{N}\frac{\alpha}{8\pi}\frac{a}{v_a}F_{\mu\nu}\tilde{F}^{\mu\nu},
\end{equation}
where $\alpha\equiv\frac{e^2}{4\pi}$, one can derive the equation of motion of photons in the presence of the axion background field,
\begin{equation}
    \partial_\nu F^{\mu\nu}=\frac{E_{\mathcal{A}}}{N}\frac{\alpha}{2\pi}\frac{a}{v_a}\partial_\nu\tilde{F}^{\mu\nu}.
\end{equation}
By using the electric and magnetic field ${\bf E}$ and ${\bf B}$, where their components are given by $F^{0i}=-E^i$, $F^{ij}=\epsilon^{ijk}B_k$, $i,\,j,\,k=1,\,2,\,3$, one can decompose the above equation into two modified Maxwell equations, 
\begin{equation} 
    \nabla\cdot\left[{\bf E}-\frac{E_{\mathcal{A}}}{N}\frac{\alpha}{2\pi}\frac{a}{v_a}{\bf B}\right]=0,
\end{equation}
\begin{equation}
\label{eq:curlB}
  \nabla\times\left[{\bf B}+\frac{E_{\mathcal{A}}}{N}\frac{\alpha}{2\pi}\frac{a}{v_a}{\bf E}\right]-\frac{\partial}{\partial t}\left[{\bf E}-\frac{E_{\mathcal{A}}}{N}\frac{\alpha}{2\pi}\frac{a}{v_a}{\bf B}\right]=0.
\end{equation}

Now, let us use \cref{eq:curlB} to find the current inflow onto the axion string. First, the axion string configuration aligned with the $z$ axis is parametrized in the cylindrical coordinates as $\Phi(\rho,\varphi,z)=\frac{v_a}{\sqrt{2}}f(\rho)e^{i\varphi}$, with the winding number $+1$. When an external electric field is applied along the string direction ${\bf E}_0=E_0\hat{\bf z}$, the non-zero gradient of $\varphi=\frac{a}{v_a}$ around the axion string gives non-zero $\nabla\times(\varphi{\bf E}_0)$, which induces the Hall current propagating in a perpendicular direction to the axion string,
\begin{equation}
    {\bf j}_{\rm in}(\rho)=\nabla\times{\bf B}_{\rm in}=-\frac{E_{\mathcal{A}}}{N}\frac{\alpha}{2\pi}\nabla\times(\varphi{\bf E}_0)=-\frac{E_{\mathcal{A}}}{N}\frac{\alpha}{2\pi}\frac{E_0}{\rho}\hat{\boldsymbol\rho}.
\end{equation}
Outside the string, the current is conserved,
\begin{equation}
    \nabla \cdot  {\bf j}_{\rm in}(\rho>0) = \nabla\cdot(\nabla\times{\bf B}_{\rm in})=0,
\end{equation}
\begin{equation}
    \partial_t \rho_{\rm in}(r>0)=\partial_t  \nabla\cdot\left[\frac{E_{\mathcal{A}}}{N}\frac{\alpha}{2\pi}\varphi{\bf B}\right]=-\frac{E_{\mathcal{A}}}{N}\frac{\alpha}{2\pi}\varphi\nabla\cdot(\nabla\times {\bf E}_0)=0.
\end{equation}
On the string, the divergence of the current is given by,
\begin{equation}
\label{eq:j_inflow}
\nabla\cdot  {\bf j}_{\rm in}(\rho)=-\frac{E_{\mathcal{A}}}{N}\frac{\alpha}{2\pi}\nabla\cdot\left(\nabla\times(\varphi{\bf E}_0)\right)=-\frac{E_{\mathcal{A}}}{N}\frac{\alpha}{2\pi}\left(\nabla\times\nabla\varphi\right)\cdot{\bf E}_0=-\frac{E_{\mathcal{A}}}{N}\alpha\delta(x)\delta(y)E_0,
\end{equation}
where we used $[\partial_x,\partial_y]\varphi=2\pi\delta(x)\delta(y)$ to get the last identity, and $\delta(x)\delta(y)$ is the 2D Dirac-delta function on the $x$-$y$ plane. So, the Hall current is present as long as there is an external electric field present around an axion string.

\section{S-matrix and phase-space integral of the zero-mode decay\label{app:s-matrix}}

In this section, we discuss the $3+1$\,D formalism, which is the standard calculation of S-matrix, where the initial and final states are chosen to be the asymptotic states far away from the interaction point. The idea is to insert the field operators into the S-matrix and then carry out the $d^4x$ integral by separating the integral by the coordinate variables in the longitudinal and the transverse direction of the string. We will show this calculation in a circular string loop configuration and use the coordinates $\{r,\theta,\rho,\varphi\}$ given in \cref{fig:coord-loop}.

By using the zero-mode field operator, we can directly construct the S-matrix for a zero-mode initial state decaying to N-particle final states.
\begin{equation}
   i \hat{T}=\lambda_D\langle p_{f1} p_{f2} ... p_{fN}|\int d^4x T\{\mathcal{O}((\phi_1...\bar{\psi})_{\rm SM}\psi^{(0)}_R)\}|p \rangle,
\end{equation}
where $\lambda_D\mathcal{O}((\phi_1...\bar{\psi})_{\rm SM}\psi^0_R)$ represent the decay operator. The calculation of the S-matrix is then straightforward. First, the field operators act on the initial and final states, which gives the corresponding wavefunctions; the spinor product can be factored out by now which can be later added back and evaluated when doing the spin sum, and we can only focus on the scalar component here:
\begin{equation}
   i \hat{T}_{\rm scalar}=\lambda_D\int d^4x \mathcal{F}(\rho)e^{i{\bf p}_f\cdot {\bf x}}e^{-ip(t-R\theta)},
\end{equation}
where $p_f^\mu=p_{f1}^\mu+...+p_{fN}^\mu=(E_f,{\bf p}_f)$ is the total 4-momentum of the final states, and $p=E$ is the initial energy or momentum of a zero mode. The $d^4x$-integral can be performed in the following orders: 

$\bullet$ 1. Integrate over $dt$ gives the Dirac-delta function resulting from the energy conservation,
\begin{equation}
    \int dt e^{i(E_f-E)t} = (2\pi)\delta(E_f-E).
\end{equation}

$\bullet$ 2. Integrate over $d\theta$ gives the overlap between the final-state wavefunctions and the zero-mode wavefunction in the longitudinal direction (or the zero-mode propagation direction).

First, one needs to use partial wave decomposition to rewrite the final-state plane wave in cylindrical coordinates,
\begin{equation}
    e^{-i{\bf p}_f\cdot {\bf x}}=e^{-i{\bf p}_{fP}\cdot \boldsymbol{r}}e^{-ip_{f,z}z}=\sum_{m=-\infty}^{+\infty}(-i)^mJ_m(|{\bf p}_{fP}|r)e^{-im(\theta-\theta_f)}e^{-ip_{f,z}z},
\end{equation}
where ${\bf p}_{fP}\equiv {\bf p}_{f,x}+{\bf p}_{f,y}$ is the planar component of ${\bf p}_f$, and it satisfies ${\bf p}_f^2={\bf p}_{fP}^2+p_{f,z}^2$, ${\boldsymbol{r}}=(x,y)$ is the polar radial vector, $r=|\boldsymbol{r}|$, $\theta_f=\tan^{-1}(p_{f,y}/p_{f,x})$, and $J_m(x)$ is the Bessel function with order $m$.

Now, one can use the identity 
\begin{equation}
    \int_0^{2\pi} d\theta e^{i(m-n)\theta}=(2\pi)\delta_{mn}
\end{equation}
to evaluate the $\theta$-integral, and setting $n\equiv ER$ which is an integer so that the zero-mode wavefunction can satisfy the periodic boundary condition under $\theta\rightarrow\theta+2\pi$,
\begin{equation}
    \int_0^{2\pi} d\theta \sum_{m=-\infty}^{+\infty}(-i)^mJ_m(|{\bf p}_{fP}|r)e^{i(n-m)\varphi}e^{im\varphi_f} = (2\pi)(-i)^ne^{in\varphi_f}J_n(|{\bf p}_{fP}|\rho).
\end{equation}

$\bullet$ 3. Finally, integrate the two transverse directions $(\hat{r},\hat{z})$ with respect to the zero mode propagation direction $\hat{\theta}$. 

To do that, one can rewrite the original coordinates $(r,z)$ to the local polar coordinates,
\begin{equation}
    r=R+\rho\cos\varphi,~~z=\rho\sin\varphi.
\end{equation}
The Jacobian determinant is $dr dz=d\rho d\varphi\rho$. The integral in the local polar coordinates is 
\begin{equation}
    I_T=\int d\rho d\varphi\rho(R+\rho\cos\varphi)\mathcal{F}(\rho)J_n(|{\bf p}_{fP}|(R+\rho\cos\varphi))e^{-ip_{f,z}\rho\sin\varphi}.
\end{equation}
Since we are interested in the large $R$ limit ($R\sim Q/v_a\gg\delta$), we can approximate and perform the integral analytically. We exploit the saddle point of the zero-mode transverse wavefunction $\mathcal{F}(\rho)$ which forces the range of $\rho$ within the spread of the zero-mode wavefunction $m_\psi^{-1}\sim\delta$ (suppose $y_\psi\sim1$), so we can take the leading order term of the integral up to the accuracy of $\mathcal{O}(\delta/R)$:
\begin{equation}
    I_{T0}= RJ_n(|{\bf p}_{fP}|R)\int d\rho \rho d\varphi \mathcal{F}(\rho)e^{-ip_{f,z}\rho\sin\varphi}\approx  Rm_\psi^{-1}J_n(|{\bf p}_{fP}|R)J_0(p_{f,z}m_\psi^{-1}) .
\end{equation}
Since the zero modes in the string can fill up to the ``Fermi surface'', $k_F\sim m_\psi\sqrt{m_\psi R}$ (see \cref{{sec:qt}}), the integer $n$ can be quite large, so it is natural to consider the large order asymptotic form of the Bessel function, which is given by
\begin{equation}
    J_\nu(\nu\,{\rm sech} \alpha)=\frac{e^{\nu({\rm tanh}\alpha-\alpha)}}{\sqrt{2\pi\nu\, {\rm tanh}\alpha}}\sum_{m=0}^\infty\frac{\Gamma(m+\frac{1}{2})}{\Gamma(\frac{1}{2})}\frac{A_m}{(\frac{1}{2}\nu\, {\rm tanh}\alpha)^m},
\end{equation}
where $A_0=1,\,A_1=\frac{1}{8}-\frac{5}{24}{\rm coth}^2\alpha,\,...$. We can take only the leading term $m=0$ and match $\nu=n$, $\nu\,{\rm sech} \alpha=|{\bf p}_{fP}|R=n|{\bf p}_{fP}|/E$, ${\rm tanh}\alpha=\sqrt{1-{\rm sech}^2\alpha}=\sqrt{1-\left(\frac{|{\bf p}_{fP}|}{E}\right)^2}$,
\begin{eqnarray}
    J_n(|{\bf p}_{fP}|R)&\approx& \frac{1}{\sqrt{2\pi R\sqrt{E^2-|{\bf p}_{fP}|^2}}}e^{n\left(\sqrt{1-\left(\frac{|{\bf p}_{fP}|}{E}\right)^2}-{\rm tanh}^{-1}\sqrt{1-\left(\frac{|{\bf p}_{fP}|}{E}\right)^2}\right)}\nonumber\\
    &\approx&\frac{1}{\sqrt{2\pi R\sqrt{E^2-|{\bf p}_{fP}|^2}}}e^{-\frac{n}{3}\left(1-\left(\frac{|{\bf p}_{fP}|}{E}\right)^2\right)^{3/2}}.
\end{eqnarray}

$\bullet$ 4. In summary, the S-matrix is 
\begin{equation}
    i\hat{T} = i\hat{T}_{\rm scalar} \times \text{spinor product},
\end{equation}
where 
\begin{equation}
\label{eq:T_scalar}
    i\hat{T}_{\rm scalar}=\lambda_D(2\pi)\delta(E_f-E)(-i)^ne^{in\theta_f}\sqrt{2\pi R}m_\psi^{-1}J_0(p_{f,z}m_\psi^{-1})\frac{e^{-\frac{n}{3}\left(1-\left(\frac{|{\bf p}_{fP}|}{E}\right)^2\right)^{3/2}}}{\sqrt{\sqrt{E^2-|{\bf p}_{fP}|^2}}},
\end{equation}
Therefore, the decay rate is 
\begin{eqnarray}
\Gamma(E)&=&\prod_{\text{LIPS}}\sum_{\rm spin} (\text{spinor product})\frac{|\hat{T}_{\rm scalar}|^2}{\Delta t \,2\pi R}\\
&=&\prod_{\text{LIPS}}\sum_{\rm spin} (\text{spinor products})\lambda_D^2(2\pi)\delta(E_f-E)m_\psi^{-2}J_0^2(p_{f,z}m_\psi^{-1})\frac{e^{-\frac{2n}{3}\left(1-\left(\frac{|{\bf p}_{fP}|}{E}\right)^2\right)^{3/2}}}{\sqrt{E^2-|{\bf p}_{fP}|^2}},\nonumber
\end{eqnarray}
where $\Delta t =(2\pi)\delta(E)|_{E=0}$ is the time interval, $\prod_{\rm LIPS}=\int d^{3N}p_{f1}p_{f2}...p_{fN}/(2\pi)^{3N}/(2E_{f1}2E_{f2}...2E_{fN})$ is Lorentz-invariant phase space.

The decay rate can be further simplified to a closed form when considering the limit $n=ER\rightarrow \infty$. In the following, we will consider the 2-particle final states and the result can be generalized to the N-particle final states.

\paragraph{Massless two-body decay}
Suppose the 2-particle final states are $q$ (denoted as particle 1) and $h$ (denoted as particle 2), which gives the spin-summed result  
\begin{eqnarray}
\label{eq:spinsum_2}
    \sum_{\rm spin}     \left(\bar{u}_q(p_{f1})\eta_R\right)\left(\bar{\eta}_Ru_q(p_{f1})\right)&=&
\sum_s(\sqrt{p_{f1}\cdot\bar{\sigma}}{\xi^\dagger}^s,\sqrt{p_{f1}\cdot\sigma}{\xi^\dagger}^s)
    \left(\begin{matrix}
        0 & 0\\
        A & 0\\
    \end{matrix}\right)
     \left(\begin{matrix}
        \sqrt{p_{f1}\cdot\sigma}{\xi}^s\\
        \sqrt{p_{f1}\cdot\bar{\sigma}}\xi^s\\
    \end{matrix}\right),\nonumber\\
    &=&\text{Tr}[(p_{f1}\cdot\sigma)A] \nonumber\\
    &=&\frac{1}{2}(E_{f1}-p_{1x}\sin\theta+p_{1y}\cos\theta)\nonumber\\
    &=& \frac{1}{2}p\cdot p_{f1} /E,
\end{eqnarray}
where $\eta_R\equiv \frac{1+\gamma_5}{2}\eta$ is the right-handed Dirac spinor of zero modes, which is given by \cref{eq:zm_solution_circular}, $A=\frac{1}{4}\left(\begin{matrix}
    1 & ie^{-i\theta} \\ -ie^{i\theta} & 1\\
\end{matrix}\right)$ is the non-zero block matrix of $\eta_R\bar{\eta}_R$, and $\xi_i~(i=1,2)$ are the two spin eigenstates of the quark.
The $\theta$ dependence in eq.\,(\ref{eq:spinsum_2}) indicates we should integrate it in S-matrix before doing the spin sum. This results in modifying the order of the Bessel function by unit 1. However, in practice, since $n=ER\gg1$, $J_{n\pm 1}\approx J_n$, we thus approximate the Bessel function as unchanged.

The phase space integral is
\begin{equation}
\label{eq:I_ps_loop_app}
    I_{\rm PS}=\int \frac{d^{3}p_{f1}}{(2\pi)^3(2E_{f1})}\frac{d^{3}p_{f2}}{(2\pi)^{3}(2E_{f2})} (2\pi)\delta(E_f-E)J_0^2(p_{f,z}m_\psi^{-1})E_{f1}\frac{e^{-\frac{2n}{3}\left(1-\left(\frac{|{\bf p}_{fP}|}{E}\right)^2\right)^{3/2}}}{\sqrt{E^2-|{\bf p}_{fP}|^2}}.
\end{equation}

$\bullet$ 1. Decompose the momentum variables  in the spherical coordinates, 
\begin{equation}
    d^3{\bf p}_{i}=p_ip_{i}dp_{i}d\cos\theta_id\varphi_{i}=p_iE_idE_id^2\Omega_i.
\end{equation}

$\bullet$ 2. Define $E_-\equiv E_1-E_2$, and the change of variables gives
\begin{equation}
    dE_1dE_2 = \frac{1}{2}dE_fdE_-.
\end{equation}
The $E_f$-integral can be thus directly evaluated with the help of $\delta(E_f-E)$, which sets $E_f=E$ in the integrand.

$\bullet$ 3. Define two dimensionless quantities normalized by the total energy,
\begin{equation}
    x_{-}\equiv E_{-}/E,~ {\bf x}_{fP}\equiv {\bf p}_{fP}/E,
\end{equation}
and express the two momenta $p_1,\,p_2$, and ${\bf x}_{fP}^2$ in terms of $x_-$:
\begin{equation}
    p_1=E_1=\frac{E_f+E_-}{2}=\frac{E(1+x_-)}{2}, ~p_2=E_2=\frac{E_f-E_-}{2}=\frac{E(1-x_-)}{2},
\end{equation}
\begin{equation}
    {\bf x}^2_{fP} = \frac{1}{4}\left[(1+x_-)^2\sin^2\theta_1+(1-x_-)^2\sin^2\theta_2+2(1-x_-^2)\sin\theta_1\sin\theta_2\cos(\varphi_1-\varphi_2)\right].
\end{equation}

$\bullet$ 4. Use the exponential function $e^{-\frac{2n}{3}\left(1-{\bf x}_{fP}^2\right)^{3/2}}$ to determine the relevant integration range for all integration variables ${\bf y}=(x_-,\cos\theta_1,\cos\theta_2,\varphi_1,\varphi_2)$. Also, by setting $m_1=m_2=0$ (massless), we identify the most dominant contribution of the integrand is obtained near the saddle points ${\bf y}_0=(\cos\theta_1=\cos\theta_2=0,\,\varphi_1=\varphi_2)$ such that ${\bf x}^2_{fP,{\rm max}}={\bf x}^2_{fP}({\bf y}_0)=1$. Physically, this means the two momentum vectors are paralleled and confined on the $x-y$ plane. It is therefore valid to Taylor expand $g({\bf y})\equiv {\bf x}_{fP}^2$ around $g({\bf y}_0)=1$:
\begin{equation}
    1-g({\bf y})=g({\bf y}_0)-g({\bf y})\approx -\frac{1}{2}\frac{\partial^2 g}{\partial {\bf y}^2}\bigg\vert_{{\bf y}_0}({\bf y}-{\bf y}_0)^2\lesssim n^{-2/3},
\end{equation}
where the first-order derivative of $g({\bf y})$ is zero (by the definition of saddle points) and the inequality is understood by substituting the Taylor expansion into the exponential function $e^{-\frac{2n}{3}\left(1-g({\bf y})\right)^{3/2}}$ and setting the exponent to be $\mathcal{O}(1)$. So, the half integration range of $y_i=(\cos\theta_1,\cos\theta_2,\varphi_-=\varphi_1-\varphi_2)$ is derived:
\begin{equation}
    \Delta y_i \equiv |y_i-y_{i,0}| =\sqrt{2}|\partial^2_{y_i}g|_{{\bf y}_0}^{-1/2}\left(\frac{2n}{3}\right)^{-1/3}.
\end{equation}
Explicitly, we get $\Delta\cos\theta_1=\sqrt{2(1+x_-)^{-1}}\left(\frac{2n}{3}\right)^{-1/3}$, $\Delta\cos\theta_2=\sqrt{2(1-x_-)^{-1}}\left(\frac{2n}{3}\right)^{-1/3}$, and $\Delta\varphi_-=\sqrt{2(1-x_-^2)^{-1}}\left(\frac{2n}{3}\right)^{-1/3}$.

\begin{figure}
    \centering
    \includegraphics[width=0.8\textwidth]{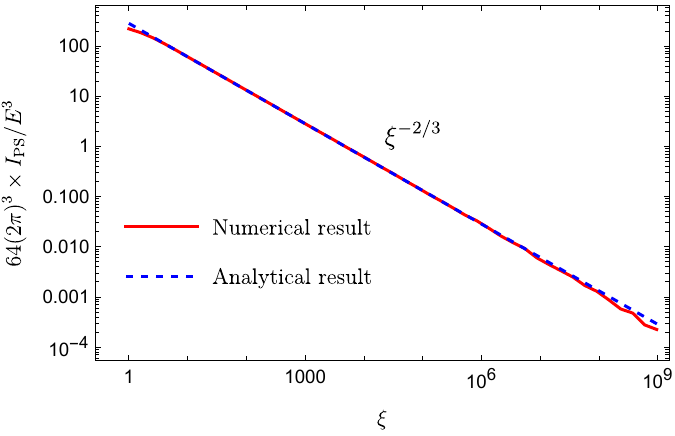}
    \caption{The power-law behavior of decay into massless two-body final states. The red solid line shows the numerical integration result while the blue dashed line shows the analytical estimation using \cref{eq:I_ps_2b_massless}, where $\xi\equiv\frac{2n}{3}$. }
    \label{fig:2body-massless-decay}
\end{figure}

The integration can thus be analytically evaluated,
\begin{equation}
\label{eq:I_ps_2b_massless}
     I_{\rm PS}=\frac{E^3}{64(2\pi)^5}\int_{-1}^{1}dx_- (1+x_-) (2\pi) 2^3 2^{3/2} \left(\frac{2n}{3}\right)^{-2/3} =\frac{\sqrt{2}E^3}{2(2\pi)^4}\left(\frac{2n}{3}\right)^{-2/3}.
\end{equation}
The result is numerically checked in \cref{fig:2body-massless-decay}. The decay rate is 
\begin{equation}
    \Gamma(E)=\frac{3^{2/3}|y_D|^2E}{2^{7/6}(2\pi)^4}\left(\frac{E}{m_\psi}\right)^2(ER)^{-2/3}.
\end{equation}

\paragraph{Massive two-body decay}
Now we consider massive final states, i.e., $m_1,\,m_2\neq0$. First, we use the spherical coordinates to express ${\bf x}_{fP}^2$:
\begin{eqnarray}
    {\bf x}_{fP}^2&=&\frac{1}{4}\left(\sqrt{(1+x_-)^2-4\frac{m_1^2}{E^2}}\sin\theta_1\cos\varphi_1+\sqrt{(1-x_-)^2-4\frac{m_2^2}{E^2}}\sin\theta_2\cos\varphi_2\right)^2\nonumber\\
    &+&\frac{1}{4}\left(\sqrt{(1+x_-)^2-4\frac{m_1^2}{E^2}}\sin\theta_1\sin\varphi_1+\sqrt{(1-x_-)^2-4\frac{m_2^2}{E^2}}\sin\theta_2\sin\varphi_2\right)^2,
\end{eqnarray}
where $x_-\equiv E_-/E=(E_1-E_2)/E$. The ${\bf x}_{fP}^2$ reaches its maximum ${\bf x}_{fP,{\rm max}}^2=1-(m_1+m_2)^2/E^2$, when $\theta_1=\theta_2=\pi/2$, $\varphi_1=\varphi_2$, and $x_-=\gamma(m_1-m_2)/E=(m_1-m_2)/(m_1+m_2)$. Therefore, we can Taylor expand $g({\bf y})\equiv{\bf x}_{fP}^2$, ${\bf y}=(x_-,\cos\theta_1,\cos\theta_2,\varphi_1,\varphi_2)$ around the saddle point and find the leading derivation terms near the $g({\bf y}_0)={\bf x}_{fP,{\rm max}}^2$:
\begin{equation}
g({\bf y})\equiv{\bf x}_{fP}^2=g({\bf y}_0)  + \frac{1}{2}\frac{d^2 g}{d{\bf y}^2}\bigg\vert_{{\bf y}_0} ({\bf y}-{\bf y}_0)^2+\mathcal{O}({\bf y}-{\bf y}_0)^3, 
\end{equation}
where $\frac{d g}{d{\bf y}}\vert_{{\bf y}_0}=0$ by the definition of a saddle point, and ${\bf y}_0=((m_1-m_2)/(m_1+m_2),0,0,\varphi,\varphi)$. To get the power-law behavior of the suppression, we take the $x_-$ direction for example. The second-order derivative to $x_-$ at the saddle point is
\begin{equation}
    \frac{\partial^2 g}{\partial x_-^2}\bigg\vert_{{\bf y}_0}=\frac{(m_1+m_2)^4}{2m_1m_2\left((m_1+m_2)^2-E^2\right)}.
\end{equation}
The derivation from the maximum to the leading order is 
\begin{equation}
    g({\bf y})-g({\bf y}_0)\approx\frac{1}{2} \frac{\partial^2 g}{\partial x_-^2}\bigg\vert_{{\bf y}_0}\left(x_--\frac{m_1-m_2}{m_1+m_2}\right)^2.
\end{equation}
Substitute it into the exponent of $\exp(-\frac{2n}{3}(1- g({\bf y}))^{3/2})$ and further Taylor expand near the saddle point, by considering the large $n$ limit where $\left(\frac{2n}{3}\right) (m_1+m_2)^3/E^3\gg 1$,\footnote{In the opposite limit, $\frac{2n}{3}(m_1+m_2)^3/E^3\ll 1$, the exponent will be close to 0, which gives a completely different scaling. For example, we will have $\frac{1}{2}|\partial^2_{x_-}g|_{{\bf y}_0}(x_--x_{-,0})^2<\left(\frac{2n}{3}\right)^{-2/3}$. This will be discussed in the small $n$ limit.}
\begin{eqnarray}
    \exp\left(-\frac{2n}{3}(1- g({\bf y}))^{3/2}\right)&=&\exp\left(-\frac{2n}{3}(1- g({\bf y}_0)+g({\bf y}_0)-g({\bf y}))^{3/2}\right)\nonumber\\
    &=&\exp\left(-\frac{2n}{3}\left(\frac{(m_1+m_2)^2}{E^2}+g({\bf y}_0)-g({\bf y})\right)^{3/2}\right)\nonumber\\
    &=&\exp\left(-\frac{2n}{3}\left(\frac{m_1+m_2}{E}\right)^3\right)\left[1+n\frac{m_1+m_2}{E}(g({\bf y})-g({\bf y}_0))\right].~~~~~~~~~
\end{eqnarray}
Note that the terms in the square bracket set the bound of the integration variables around the saddle point. Here, for our example, we get $\frac{1}{2} |\frac{\partial^2 g}{\partial x_-^2}|_{{\bf y}_0}(x_--x_{-,0})^2\le \frac{2}{3}\frac{E}{m_1+m_2}\left(\frac{2n}{3}\right)^{-1}$. Similarly, we find $\frac{1}{2} |\frac{\partial^2 g}{\partial \cos\theta_i^2}|_{{\bf y}_0}(\cos\theta_i-\cos\theta_{i,0})^2\le \frac{2}{3}\frac{E}{m_1+m_2}\left(\frac{2n}{3}\right)^{-1}$ and $\frac{1}{2} (|\frac{\partial^2 g}{\partial \varphi_1^2}|_{{\bf y}_0}+|\frac{\partial^2 g}{\partial \varphi_2^2}|_{{\bf y}_0})(\varphi_1-\varphi_2)^2\equiv\frac{1}{2} (|\frac{\partial^2 g}{\partial \varphi_1^2}|_{{\bf y}_0}+|\frac{\partial^2 g}{\partial \varphi_2^2}|_{{\bf y}_0})\varphi_-^2 \le \frac{2}{3}\frac{E}{m_1+m_2}\left(\frac{2n}{3}\right)^{-1}$.
We define $\Delta x_-\equiv |x_--x_{-,0}|$, $\Delta \cos\theta_i\equiv |\cos\theta_i-\cos\theta_{i,0}|$, and $\Delta\varphi_-=|\varphi_1-\varphi_2|$ as half of the integration range.

Now, the phase space integral in spherical coordinates $d^3{\bf p}_i=p_iE_idE_id\cos\theta_i d\varphi_i$, with $dE_1dE_2=\frac{1}{2}dE_fdE_-$, $p_i=\sqrt{E_i^2-m_i^2}$, $i=1,\,2$, becomes
\begin{eqnarray}
     I_{\rm PS}&=&\int \frac{d^{3}{\bf p}_{1}}{(2\pi)^{3}(2E_{1})}\frac{d^{3}{\bf p}_{2}}{(2\pi)^{3}(2E_{2})} (2\pi)\delta(E_f-E)E_{1}\frac{e^{-\frac{2n}{3}\left(1-g({\bf y})\right)^{3/2}}}{E\sqrt{1-g({\bf y})}}\nonumber\\
     &=&\frac{E^{-1}}{8(2\pi)^5}\int dE_f\delta(E_f-E)dE_-d\cos\theta_1 d\varphi_1d\cos\theta_2 d\varphi_2 p_1p_2\frac{E_f+E_-}{2}\frac{e^{-\frac{2n}{3}\left(1-g({\bf y})\right)^{3/2}}}{\sqrt{1-g({\bf y})}}\nonumber\\
     &=&\frac{E^3}{64(2\pi)^5}\int dx_-d\Omega_1d\Omega_2(1+x_-)\sqrt{(1+x_-)^2-4\frac{m_1^2}{E^2}}\sqrt{(1-x_-)^2-4\frac{m_2^2}{E^2}}\frac{e^{-\frac{2n}{3}\left(1-g({\bf y})\right)^{3/2}}}{\sqrt{1-g({\bf y})}}\nonumber\\
     &\approx&\frac{E^3}{64(2\pi)^5}\int dx_-d\Omega_1 d\Omega_2(1+x_-)\sqrt{(1+x_-)^2-4\frac{m_1^2}{E^2}}\sqrt{(1-x_-)^2-4\frac{m_2^2}{E^2}}\frac{E}{m_1+m_2}\nonumber\\
     &\times&\exp\left(-\frac{2n}{3}\left(\frac{m_1+m_2}{E}\right)^3\right)\left[1+n\frac{m_1+m_2}{E}(g({\bf y})-g({\bf y}_0))\right]\nonumber\\
      &\approx& \frac{4E^3}{64(2\pi)^5}\frac{E}{m_1+m_2}\exp\left(-\frac{2n}{3}\left(\frac{m_1+m_2}{E}\right)^3\right)2^4(2\pi)\Delta x_-^2\Delta\cos\theta_1 \Delta\cos\theta_2\Delta\varphi_-\nonumber\\
     &=&\frac{3E^3}{8(2\pi)^4}\exp\left(-\frac{2n}{3}\left(\frac{m_1+m_2}{E}\right)^3\right)\left(\frac{4}{3}\frac{E}{m_1+m_2}\right)^{7/2}\left(\frac{2n}{3}\right)^{-5/2}\frac{E^2}{(m_1+m_2)^2}.
\end{eqnarray}
where we approximate $\sqrt{(1-x_-)^2-4\frac{m_2^2}{E^2}}\approx(1-x_-)$, $\sqrt{(1+x_-)^2-4\frac{m_1^2}{E^2}}\approx(1+x_-)\approx 2$, we only keep the leading-order term in the square bracket, and we use the following second-order derivatives for each variable: $\partial^2_{\cos\theta_1}g\vert_{{\bf y}_0}=-2m_1/(m_1+m_2)$, $\partial^2_{\cos\theta_2}g\vert_{{\bf y}_0}=-2m_2/(m_1+m_2)$, and $\partial^2_{\varphi_1}g\vert_{{\bf y}_0}=\partial^2_{\varphi_2}g\vert_{{\bf y}_0}=-2m_1m_2/(m_1+m_2)^2$.

\begin{figure}
    \centering
    \includegraphics[width=0.8\textwidth]{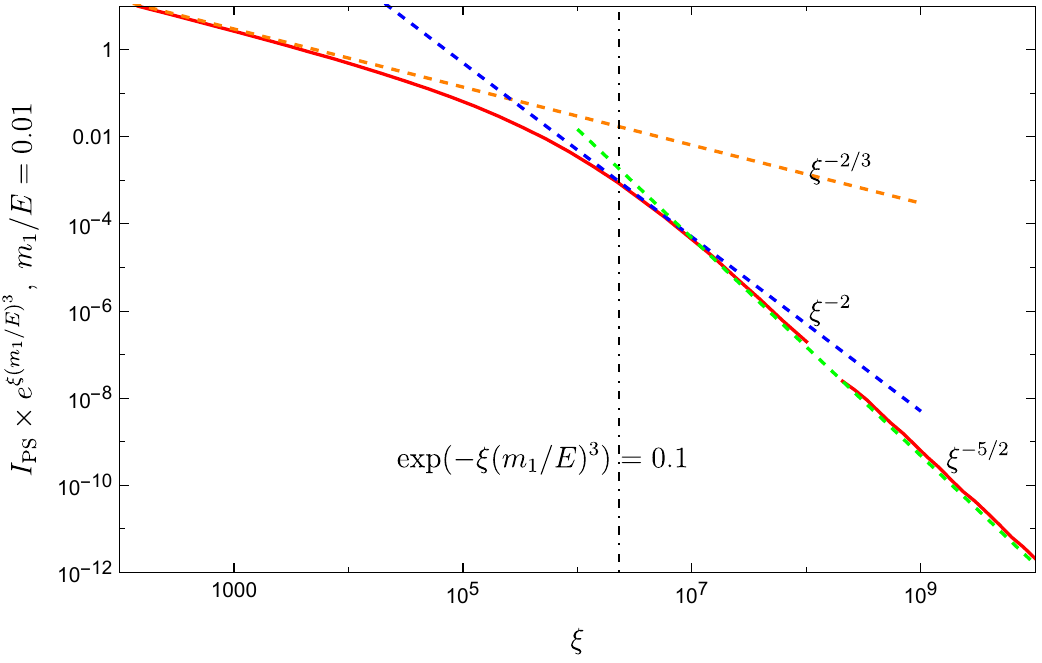}
    \caption{The power-law behavior of decay into massive two-body final states. We choose the mass of two final-state particles, $m_1=0.01E$ and $m_2=10^{-12}E$, where $\xi\equiv \frac{2n}{3}$. The red solid line shows the numerical integration result while the dashed lines show three different power-law fitting.}
    \label{fig:2body-massive-decay}
\end{figure}

The asymptotic decay rate is
\begin{equation}
\label{eq:Gamma_2b_massive}
    \Gamma(E)=\frac{2^{3/2}|y_D|^2E}{(2\pi)^4}\frac{E^2}{m_\psi^2}\exp\left(-\xi\left(\frac{m_1+m_2}{E}\right)^3\right)\left(\frac{E}{m_1+m_2}\right)^{11/2}(ER)^{-5/2}.
\end{equation}
We show the numerical result of using the Monte Carlo integrator (vegas) in \cref{fig:2body-massive-decay}. The phase space integral has the exact power-law suppression $n^{-5/2}$ or $(ER)^{-5/2}$ when $\frac{2n}{3}\gg E^3/(m_1+m_2)^3\approx 10^6$ as the analytical result \cref{eq:Gamma_2b_massive} predicted.

We also check the small $n$ limit. The exponent becomes 
\begin{equation}
    -\left(\left[\frac{2n}{3}\left(\frac{m_1+m_2}{E}\right)^3\right]^{2/3}+\left(\frac{2n}{3}\right)^{2/3}[g({\bf y}_0)-g({\bf y})]\right)^{3/2}.
\end{equation}
Since the first term is much smaller than the second term, we get the scaling of the integration variables: $\frac{1}{2} |\frac{\partial^2 g}{\partial x_-^2}|_{{\bf y}_0}(x_--x_{-,0})^2\le \left(\frac{2n}{3}\right)^{-2/3}$. Similarly, we find $\frac{1}{2} |\frac{\partial^2 g}{\partial \cos\theta_i^2}|_{{\bf y}_0}(\cos\theta_i-\cos\theta_{i,0})^2\le \left(\frac{2n}{3}\right)^{-2/3}$ and $\frac{1}{2} (|\frac{\partial^2 g}{\partial \varphi_1^2}|_{{\bf y}_0}+|\frac{\partial^2 g}{\partial \varphi_2^2}|_{{\bf y}_0})(\varphi_1-\varphi_2)^2\equiv\frac{1}{2} (|\frac{\partial^2 g}{\partial \varphi_1^2}|_{{\bf y}_0}+|\frac{\partial^2 g}{\partial \varphi_2^2}|_{{\bf y}_0})\varphi_-^2 \le \left(\frac{2n}{3}\right)^{-2/3}$. Substitute the corresponding second-order derivatives, we can find the corresponding integration limits:
\begin{eqnarray}
   &~& \Delta x_- = \sqrt{\frac{4m_1m_2E^2}{(m_1+m_2)^4}}\left(\frac{2n}{3}\right)^{-1/3},~\Delta\cos\theta_1 = \sqrt{\frac{m_1+m_2}{m_1}}\left(\frac{2n}{3}\right)^{-1/3},\nonumber\\
    &~&\Delta\cos\theta_2 = \sqrt{\frac{m_1+m_2}{m_2}}\left(\frac{2n}{3}\right)^{-1/3},~\Delta\varphi_-=\sqrt{\frac{(m_1+m_2)^2}{2m_1m_2}}\left(\frac{2n}{3}\right)^{-1/3}.
\end{eqnarray}
Note that we need to compare the above integration ranges with their physically available ranges, i.e., $x_-\in[2\frac{m_1}{E}-1,1-2\frac{m_2}{E}]$, $\cos\theta_i\in[-1,1]$, and $\varphi_-\in[0,2\pi]$. Numerically, we find that only the integration ranges of $x_-$ and $\cos\theta_1$ are suppressed by small $n$ for our chosen parameters $\frac{m_1}{E}=0.01$ and $\frac{m_2}{E}=10^{-12}$.

\section{Gaussian modulation\label{app:GaussianModulation}}
The section below summarizes the details of the calculation of the decay rate of zero modes in a straight string with a bent string segment modeled by a Gaussian modulation. We calculate the decay rate for both the process $\psi^{(0)}\rightarrow q+h$ and the non-adiabatic process $\psi^{(0)}+\delta\Phi\rightarrow\tilde{\psi}\rightarrow q+h$. The coordinates are defined in \cref{fig:coord-straight-string}.

\paragraph{Gaussian modulation ($\psi^{(0)}\rightarrow q+h$)} 

We consider a straight string along the $z$ axis with a tiny Gaussian modulation $\Delta(z)=\epsilon\delta \exp(-z^2/(2\sigma_z^2))$ towards the $x$ axis, where $\epsilon\ll 1$ is a constant such that $\Delta(z)\ll\delta$ is satisfied, i.e., the modulation is much smaller than the string core size $\delta$, and $\sigma_z$ is related to the curvature radius $R$ of the string segment by 
\begin{equation}
\label{eq:sigmaz}
   \sigma_z^2 =  \Delta(z)_{\rm max}(R-\Delta(z)_{\rm max})=\epsilon\delta( R-\epsilon\delta) ~~ \Rightarrow \sigma_z^2\simeq\epsilon\delta R.
\end{equation}
Due to the perturbativeness of the modulation, it is good enough to only consider the decay happens at the range of $z\in[-\sigma_z,\sigma_z]$ and approximate the modulation by a circle arc with an open angle $\Theta_{\rm m}=2\sigma_z/R$, where we shift the origin of the coordinate system in \cref{fig:coord-straight-string} to $\boldsymbol{\rho}=(-(R-\epsilon\delta),\,0,\,0)$ such that we can use the open angle $\Theta$ to simplify the $\zeta$-integral, similar to what we did for a circular loop configuration. By equating $\zeta=R\Theta$, the $\zeta$-integral in \cref{eq:S-matrix-ws} becomes
\begin{equation}
    I_\zeta = R\int_0^{\Theta_{\rm m}}d\Theta e^{-iR{\bf p}_{fP}\cdot \hat{\boldsymbol{\rho}}}e^{iER\Theta}=\sigma_z\sum_{n} 2(-i)^n e^{i(ER-n)\sigma_z/R}\text{sinc}((ER-n)\sigma_z/R)J_n(|{\bf p}_{fP}|R)e^{in\Theta_f},
\end{equation}
where $\text{sinc}(x)\equiv\sin(x)/x$, ${\bf p}_{fP}$ is the total final-state planar momentum on the $x-z$ plane (the same plane where the bent string segment is located), and $\Theta_f=\tan^{-1}( p_{f,z}/ p_{f,x})$. Different from the circular string loop result, $ER$ is not necessarily quantized in this case. Thus, the S-matrix result is a superposition of all wavefunction overlap $J_n(|{\bf p}_{fP}|R)$ and there is a finite overlap spreading between the zero-mode and final-state wavefunctions which is given by the $\text{sinc}((ER-n)\sigma_z/R)$ function. The S-matrix is 
\begin{equation}
    i\hat{T}=\kappa \sigma_z(2\pi)\delta(E_f-E)(\bar{u}(p_1)\eta_R)\sum_n 2(-i)^n e^{i(ER-n)\sigma_z/R}\text{sinc}((ER-n)\sigma_z/R)J_n(|{\bf p}_{fP}|R)e^{in\Theta_f}.
\end{equation}

The effective decay rate of this configuration is given by
\begin{equation}
    \Gamma(E)=\int\frac{d^3{\bf p}_1}{(2\pi)^3(2E_1)}\int\frac{d^3{\bf p}_2}{(2\pi)^3(2E_2)}\frac{|\hat{T}|^2}{\Delta t\,L_{\rm str}}=4|\kappa|^2\sigma_z I_{\rm PS},
\end{equation}
where $L_{\rm str}=\sigma_z$ and the phase space integral is
\begin{eqnarray}
\label{eq:I_ps_bend}
    I_{\rm PS}&\equiv& \int\frac{d^3{\bf p}_1}{(2\pi)^3(2E_1)}\int\frac{d^3{\bf p}_2}{(2\pi)^3(2E_2)}(2\pi)\delta(E-E_f)E_1\bigg[\sum_n  \text{sinc}^2((ER-n)\sigma_z/R)J_n^2(|{\bf p}_{fP}|R)\nonumber\\
    &+&\sum_{m>n}2\cos\left((m-n)\left(\frac{\sigma_z}{R}-\Theta_f+\frac{\pi}{2}\right)\right)\text{sinc}\left(\frac{(ER-m)\sigma_z}{R}\right)\text{sinc}\left(\frac{(ER-n)\sigma_z}{R}\right)\nonumber\\
    &\times&J_m(|{\bf p}_{fP}|R)J_n(|{\bf p}_{fP}|R)\bigg].
\end{eqnarray}
The second sum in the above equation represents the interference between different partial waves $J_m$ and $J_n$ $(m\neq n)$ due to their non-vanishing overlapping. For now, we neglect the interference terms for simplicity and only keep the leading terms from the first sum. The most dominant term is at $n_0= \text{floor}(ER)$, which is an integer closely equal to $ER$, such that $0\le(ER-n_0)<1$. So, the sinc function in eq.\,(\ref{eq:I_ps_bend}) is $\mathcal{O}(1)$ and the effective decay rate from the $J_{n_0}^2$ term is equal to the circular loop result multiplied by an additional suppression $4\frac{\sigma_z}{R}=4\sqrt{\frac{\epsilon\delta}{R}}$ due to the probability to have a Gaussian modulation in a straight string. The remaining subdominant terms are given by a range of $n\in[n_0-\Delta n,n_0+\Delta n]$, where $\Delta n\sim\text{floor}(\sigma_z^{-1}R)$ corresponding to the first peak of the $\text{sinc}(x)$ function.
By taking the large $n_0$ limit, we get the asymptotic decay rate of each $J_n^2$ term is $\propto n^{-2/3}$. By summing up all the $n$, we get the effective decay rate. 
\begin{equation}
    \Gamma(E)=\sum_{n=n_0-\Delta n}^{n_0+\Delta n}\Gamma_{n}\gtrsim  \Gamma_{n_0}= \frac{2^{5/6}3^{2/3}|y_D|^2\sqrt{\epsilon \delta E}E}{(2\pi)^4}\left(\frac{E}{m_\psi}\right)^2 (ER)^{-7/6},
\end{equation}
where we only keep the leading term $n_0= \text{floor}(ER)$ and the $n$-th decay rate contribution is
\begin{equation}
    \Gamma_n=4 \sqrt{\frac{\epsilon \delta}{R}}\text{sinc}^2\left(\frac{(ER-n)\sigma_z}{R}\right)\frac{3^{2/3}|y_D|^2E}{2^{7/6}(2\pi)^4}\left(\frac{E}{m_\psi}\right)^2 n^{-2/3}.
\end{equation}

\paragraph{Gaussian modulation (non-adiabatic)} 
For the Gaussian modulation, we set $\Delta(z)=\epsilon \delta e^{-\frac{z^2}{2\sigma_z^2}}$ in \cref{eq:I_z_non_ad}.
The $z$-integral can be obtained by completing the square and then using a Gaussian integral, 
\begin{equation}
\label{eq:I_z_Gaussian}
    I_z=\epsilon\delta e^{-\frac{\sigma_z^2}{2}(p_{f,z}-p_z)^2}(\sqrt{2\pi}\sigma_z),
\end{equation}
where the Gaussian integral is 
\begin{equation}
    \int_{-\infty}^{+\infty} dz\, e^{-\frac{\left(z+i\sigma_z^2(p_{f,z}-p_z)\right)^2}{2\sigma_z^2}}=\sqrt{2\pi}\sigma_z.
\end{equation}
Since the Gaussian modulation is extremely suppressed when $z\gg \sigma_z$, we can consider the decay only happens at the range of $-\sigma_z<z<\sigma_z$. The S-matrix then only contains a 2D integral. First, decompose the plane wave $e^{-i({\bf p}_{f}-{\bf p})\cdot {\boldsymbol{\rho}}}$ to partial waves in polar coordinate,
\begin{equation}
    I_{\rm 2}\equiv \int d^2x \mathcal{F}(\rho)\frac{d\ln f}{d\rho}\cos\varphi \sum_{n=-\infty}^{+\infty} (-i)^nJ_n(|{\bf p}_{fT}|\rho)e^{-i n(\varphi-\varphi_{T})},
\end{equation}
where ${\bf p}\cdot\boldsymbol{\rho}=0$ since zero-mode momentum is perpendicular to the transverse direction. By doing $\varphi$-integral, we fix the Bessel function order to $n=\pm1$ due to the presence of $\cos\varphi=(e^{i\varphi}+e^{-i\varphi})/2$ for the first term in the square bracket, and then we further Taylor expand the Bessel function $J_1(|{\bf p}_{fT}|\rho)$ around 0 by assuming $|{\bf p}_{fT}|\rho\sim \rho/R\rightarrow 0^+$ in the large curvature limit,
\begin{equation}
    I_{\rm 2}\simeq -i2\pi p_{f,x} \int d\rho \rho^2 \mathcal{F}(\rho) \frac{d\ln f}{d\rho},
\end{equation}
where we use $|{\bf p}_{fT}|\cos\varphi_T=p_{f,x}$.
The radial integral can be again evaluated by using the saddle-point approximation, where the zero-mode transverse wavefunction $\mathcal{F}(\rho)$ will restrict the integral to be evaluated at the saddle-point position $\rho_*\sim m_\psi^{-1}$. Up to the $\mathcal{O}(1)$ prefactor, we get 
\begin{equation}
    I_{\rm 2}\simeq-i p_{f,x}m^{-1}_{\psi}\left[\frac{m_\phi}{m_\psi}\text{sech}\left(\frac{m_\phi}{m_\psi}\right)\text{csch}\left(\frac{m_\phi}{m_\psi}\right)\right]\simeq -i p_{f,x}m^{-1}_{\psi},
\end{equation}
where we use the analytical form of the profile function $f(\rho)=\text{tanh}(m_\phi \rho)$ to calculate its first-order derivative.

The decay rate now becomes, 
\begin{equation}
    \Gamma(E)=\int\frac{d^3{\bf p}_1}{(2\pi)^3(2E_1)}\int\frac{d^3{\bf p}_2}{(2\pi)^3(2E_2)}\frac{|\hat{T}|^2}{\Delta t\,L_{\rm str}}=|y_D|^2(2\pi\epsilon^2\delta^2\sigma_z)m_\psi^{-2} I_{\rm PS},
\end{equation}
where $L_{\rm str}=\sigma_z$ and the phase space integral is changed by a Gaussian kernel,
\begin{equation}
    I_{\rm PS}\equiv \int\frac{d^3{\bf p}_1}{(2\pi)^3(2E_1)}\int\frac{d^3{\bf p}_2}{(2\pi)^3(2E_2)}(2\pi)\delta(E-E_f)(E_1+p_{1,z})(p_{1,x}+p_{2,x})^2e^{-\frac{\sigma_z^2}{2}(p_{f,z}-p_z)^2}.
\end{equation}
By using the same changes of variables, $E_1=(E_f+E_-)/2$ and $E_2=(E_f-E_-)/2$, $x\equiv E_-/E$, and the massless final state approximation $|{\bf p}_i|=E_i$, $i=1,2$, we can rewrite the phase space integral to
\begin{eqnarray}
\label{eq:I_PS_GM1}
I_{\rm PS}&=&\frac{E^6}{128(2\pi)^5}\int_{-1}^{1}dx \int d\Omega_1 d\Omega_2  (1+\cos\theta_1)(1+x)^2(1-x)\Big[(1+x)\sqrt{1-\cos^2\theta_1}\cos\varphi_1\nonumber\\
&+&(1-x)\sqrt{1-\cos^2\theta_2}\cos\varphi_2\Big]^2 \exp[-\frac{\sigma_z^2E^2}{8}\left((1+x)\cos\theta_1+(1-x)\cos\theta_2+2\right)^2].~~~~~~~~~~
\end{eqnarray}
The above integral without the prefactor, $I_{\rm PS}=\frac{E^6}{128(2\pi)^5} I$, has a closed form expression:
\begin{equation}
   I=\frac{16(4-4e^{-2y^2}+4y^2-3\sqrt{2\pi}y{\rm Erf}(\sqrt{2}y))}{y^6},  
\end{equation}
where $y\equiv\sigma_z E$, and ${\rm Erf}(x)$ is the error function and it asymptotically approaches 1 when $x\rightarrow1$ (because the error function is the cumulative distribution function of a Gaussian probability distribution function). Since $y\sim\sqrt{\epsilon\delta R}E\gg 1$ for a large curvature radius, i.e., $R\gtrsim E^{-2}/(\epsilon\delta)$, the leading-order term of the above integral is
\begin{equation}
    I\approx \frac{64}{y^4}.
\end{equation}
The decay rate is thus 
\begin{equation}
 \Gamma(E)=\frac{1}{32\pi^4}\sqrt{\epsilon}|y_D|^2\left(\frac{E}{m_\psi}\right)^{2}\left(\frac{\delta}{R}\right)^{3/2}\delta^{-1}.
\end{equation} 

\bibliographystyle{JHEP}
\bibliography{ref}
\end{document}